\documentclass[10pt, conference]{IEEEtran}

\usepackage{amsmath, amssymb, amsfonts}
\usepackage{algorithmic}
\usepackage{graphicx}
\usepackage{textcomp}
\usepackage{xcolor}
\usepackage{url}
\usepackage{xurl}
\usepackage[hidelinks]{hyperref}

\def\BibTeX{{\rm B\kern-.05em{\sc i\kern-.025em b}\kern-.08em
    T\kern-.1667em\lower.7ex\hbox{E}\kern-.125emX}}

\begin{document}

\title{Laminar: A Probe-First Scheduling Paradigm with Deterministic Runtime Survival}

\author{\IEEEauthorblockN{Zhengyan Chu}
\IEEEauthorblockA{\textit{Independent Researcher} \\
China \\ 
zchu1994@gmail.com}
}

\maketitle

\begin{abstract}
In exascale-oriented GPU clusters, rigid-topology jobs leave behind a fragmented post-landing ecology in which long-resident workloads and highly transient tasks compete for unstable residual capacity. Existing centralized, hierarchical, and local-first decentralized schedulers incur growing coordination and retry-amplification costs in this regime and typically stop their explicit responsibility at execution start, leaving runtime survival to indiscriminate host-level OOM heuristics. We present Laminar, a decentralized probe-first, execute-later scheduling paradigm that keeps hot-path control-plane work near $\mathcal{O}(1)$ through Zone-level probabilistic flow splitting, bounded in-Zone probing by persistent lightweight agents, and node-local arbitration. Laminar further introduces Airlock, a bounded node-local runtime-survival layer that converts severe memory pressure into an ordered policy of suspension, in-situ recovery, bounded secondary re-addressing, or reclamation. By enforcing priority-ordered survival under pressure, Laminar enables lifecycle-aware scheduling that preserves high-value long-resident work and operates closer to physical saturation without relying on protocol-level overcommitment.
\end{abstract}

\section{Introduction}
\label{sec:intro}

Modern AI clusters are entering a regime in which scheduling is
no longer only a placement problem, but an ecological one.
Exascale-oriented GPU deployments increasingly co-locate three
structurally different kinds of work.
\emph{Rigid-topology jobs}---such as large distributed training
runs---demand cross-node placement with strong locality constraints
and, once started, typically remain resident for long periods.
\emph{Long tasks} occupy node-local resources for durations
sufficient to shape the residual capacity seen by later arrivals.
\emph{Fine-grained tasks (F-tasks)}, including transient inference
requests, serving stages, routing invocations, and control agents,
arrive at high rates and often complete in tens of milliseconds.
These three classes do not stress the system in the same way, and
a scheduler that treats them as a homogeneous population risks
solving the wrong problem.

This paper is not about replacing the upper-layer scheduler that
handles rigid cross-node topology placement.
That layer remains necessary, but it operates at a different time
scale and under a different cost structure: rigid-topology jobs
are relatively infrequent control events, require global reasoning,
and once admitted tend to remain stable for long intervals.
The regime we target begins after such jobs have already landed.
What remains is a fragmented residual cluster: nodes whose
allocatable resource atoms are partially occupied, whose bitmap
geometry is uneven, and whose local memory headroom can change
faster than a global controller can safely reason about.
It is within this post-landing cluster ecology that Long tasks and
F-tasks must be admitted cheaply, started quickly, and kept alive
predictably under pressure.

In this regime, scheduling itself becomes a first-order systems
problem.
F-tasks generate sustained high-frequency arrival pressure; Long
tasks accumulate local occupancy and shape the residual capacity
seen by subsequent tasks; and operators increasingly push
utilization toward the high end of the stable range.
Under these conditions, per-task control work, stale decision
state, retries, and blind runtime failure handling directly affect
not only latency but the survivability of useful work.
A scheduler that is efficient only at admission, yet delegates
post-start contention to opaque host heuristics, solves only part
of the problem.

Existing scheduling paradigms expose distinct stress points here.
Centralized designs (e.g., Kubernetes, Slurm) maintain a logically
global resource view, but they pay for that visibility through
serialized coordination and lock contention: when a
topology-intensive or resource-heavy decision occupies the head of
a global queue, smaller tasks stall behind it even when fragmented
capacity exists elsewhere.
Hierarchical designs (e.g., Flux) partition control to reduce
single-point pressure, yet under high churn sibling schedulers
still act on stale, isolated ledgers; concurrent placements collide
near the leaves and corrections propagate back up the tree.
Local-first decentralized systems (e.g., Ray) improve median
responsiveness at moderate load, but when utilization is pushed
high and spillback becomes common, local misses expand into broad
redirection traffic and retry amplification.
In the operating regime we target, each architecture reaches a
variant of the same outcome: the hot scheduling path remains
coupled to a coordination mechanism whose cost grows precisely
when the cluster is most fragmented and reactive.

A common source of this difficulty is that existing paradigms
largely stop their explicit responsibility at placement and
execution start.
Resource discovery remains tightly coupled to heavyweight
coordination, while post-start runtime survival is typically left
outside the scheduler's primary semantic boundary.
Architecturally, this is understandable: once the hot path already
depends on global locking, hierarchical correction, or
network-visible retry chains, extending the same paradigm into
execution-time survival would only deepen the overhead problem.
But in next-generation AI clusters, this boundary is no longer
satisfactory.
When Long tasks and F-tasks coexist in the fragmented cluster left
behind by rigid-topology jobs, the system needs not only cheap
landing, but also a bounded way to avoid degenerating into
indiscriminate failure behavior under severe local pressure.

What is therefore needed is not merely a faster scheduler, but a
new scheduling paradigm: one that treats rigid-topology placement
as an upper-layer concern, focuses on the residual ecology beneath
it, admits Long tasks and F-tasks with near-constant hot-path
control cost, tolerates fragmented node-local capacity under
partial visibility, and extends declared scheduling order into
runtime survival when physical memory pressure becomes acute.
Critically, we treat survival priority as hierarchical: rigid-topology 
jobs must be preserved first, followed by Long tasks, and only then 
fine-grained tasks. When node-local memory pressure becomes acute, 
the goal is therefore not only to keep throughput high, but to avoid 
kernel-level OOM behavior that indiscriminately kills high-value rigid 
or Long jobs instead of expendable F-tasks.
Execution start should remain the closure point of the common-case
hot path, but not the end of responsibility in the exceptional
case.

This paper presents \textit{Laminar}, a decentralized scheduling
paradigm built around a single organizing principle: probe first,
execute later, and survive locally if possible.
Instead of binding search and execution into a single decision
boundary, Laminar lets tasks first move as lightweight control
probes that compete for node-local reservations; only after a
reservation holds and execution start is observed does the task
cross into execution.
Laminar operates under \emph{weak verification}: tenant-declared
priorities and resource masks are accepted as scheduling inputs
without online truth auditing, while their effect is structurally
constrained through bounded search, reservation expiry, patience
budgets, and conservative handling of stale or missing state.
Failed placement is treated as bounded dissipation rather than a
fault requiring heavyweight recovery.

Laminar organizes this control path into three decoupled layers.
At the cluster-entry layer, a Thermo-Economic Gateway (TEG)
performs probabilistic flow splitting using only Zone-level
aggregate state.
At the Zone layer, a per-task Decentralized Agent (DA) performs
bounded node addressing guided by the Zone Holographic
Availability Field (Z-HAF).
At the node layer, a Node Arbitrator closes admission at a
single-node boundary through a TTL-bounded reservation and
payload-pull transition.
This common path is deliberately engineered to keep per-success
control work approximately constant in the scale-out regime we
evaluate.

However, Laminar does not stop at execution start.
Once admitted, a task remains attached to a resident DA and may
later be suspended, resumed, or secondarily re-addressed under
node-local runtime control if the host enters severe physical
memory pressure.
To avoid reacting to every transient spike with migration while
also avoiding blind kernel-level failure behavior, Laminar
introduces \emph{Airlock}, a bounded runtime survival layer between
local overload and irreversible task death.
A suspended task first receives a local suspension threshold that
favors in-situ recovery; if pressure persists, its resident DA is
reactivated for bounded secondary re-addressing under a shared
survival deadline; if neither local recovery nor successful
secondary landing completes in time, both the suspended task and
its control state are reclaimed by bounded local cleanup.
Laminar thereby extends declared priority from an admission-time 
ordering principle into a bounded runtime-survival ordering principle.

The resulting scope is intentionally layered.
Cross-node gang placement for rigid-topology jobs remains above
Laminar.
Laminar instead targets the fragmented post-landing cluster beneath
that layer, where Long tasks and F-tasks must be placed with
near-$O(1)$ control-path cost and must not be abandoned to
undifferentiated failure behavior after start.
Our claim is therefore not that Laminar replaces every scheduler in
the stack, but that it fills a missing paradigm layer in
next-generation AI clusters: lifecycle-aware scheduling for the
residual ecology left behind by large rigid workloads.

\smallskip
\noindent\textbf{Summary of Contributions.}
\begin{itemize}

\item \textbf{A new scheduling scope for next-generation AI
clusters.}
We identify the post-landing cluster ecology---created after
rigid-topology jobs have been placed---as a distinct systems
problem, and argue that Long tasks and F-tasks require a scheduling
paradigm that addresses both fast admission and bounded runtime
survival.

\item \textbf{A probe-first scheduling paradigm under weak
verification.}
Laminar explicitly separates resource discovery from execution
start, treats failed placement as bounded dissipation, and avoids
strongly consistent global coordination on the hot path.

\item \textbf{Approximately constant per-success control work.}
Entry-side routing depends only on Zone summaries, DAs perform
bounded in-Zone addressing, and node-local arbitration closes
placement at a single node, yielding approximately constant
control-plane work per successful landing in the evaluated
scale-out regime.

\item \textbf{A reservation-centered execution boundary with
lifecycle continuity.}
A winning probe obtains a TTL-bounded logical reservation, and
execution is recognized only after payload pull succeeds within the
valid window; after start, the same task remains attached to a
resident DA rather than disappearing from the control model.

\item \textbf{A bounded runtime survival layer.}
Laminar introduces Airlock, which converts severe node-local memory
pressure into an ordered runtime policy---local suspension,
in-situ recovery, bounded secondary re-addressing, or bounded
reclamation---rather than blind kernel-level destruction.

\item \textbf{Comparative and mechanism-level evaluation.}
We evaluate Laminar against Slurm-like, Ray-like, and Flux-like
baselines under mixed-load, scale-out, and partial-visibility
conditions, and isolate key mechanisms to show where the design
remains stable as competing paradigms become coordination-bound,
structure-bound, or retry-bound.

\end{itemize}
\section{Background and Motivation}
\label{sec:motivation}

We now characterize the operating regime that motivates Laminar.
Our goal is not to replace the upper-layer scheduler responsible
for rigid cross-node placement, but to address the scheduling
ecology that emerges after such placement has already occurred.
In next-generation AI and exascale clusters, this distinction
matters: once large rigid-topology jobs have landed, the system
beneath them is no longer a clean pool of homogeneous free
capacity, but a fragmented residual cluster in which Long tasks
and fine-grained tasks must compete under high churn, partial
visibility, and tight physical resource margins.

\subsection{Task Ecology After Rigid-Topology Placement}

Modern AI datacenters increasingly co-locate three structurally
different classes of work.
The first class consists of \emph{rigid-topology jobs}---large
distributed training runs that demand accelerator groups spanning
multiple nodes and impose strict physical locality constraints
(e.g., adjacency within an NVLink domain or a non-blocking
RoCE fabric).
Cross-node gang placement for these jobs is resolved by
higher-layer schedulers before they enter execution; what
Laminar observes is the resulting per-node bitmap state---nodes
whose resource atoms are partially or heavily consumed, leaving
residual regions of varying size and contiguity.
These jobs are relatively infrequent control events, require
global reasoning, and once admitted tend to remain resident for
long intervals.
Rigid-topology placement is therefore a necessary upper-layer
concern, but not the one Laminar is designed to solve.

The second class consists of \emph{Long tasks}.
These tasks do not require the same cross-node gang semantics as
rigid-topology jobs, yet they occupy node-local resources for
durations long enough to shape the residual capacity seen by
future arrivals.
A Long task is therefore not merely an isolated execution event:
it is a persistent ecological object that changes local slack,
fragments available bitmap geometry, and alters the contention
surface against which later arrivals must search.

The third class consists of \emph{fine-grained tasks (F-tasks)},
including transient inference requests, serving stages,
Mixture-of-Experts (MoE) routing invocations, agentic steps, and
control-plane microservices.
F-tasks execute for only tens of milliseconds (typically
$T_{\text{exec}}$ between 10 and 50~ms), yet cluster-wide
arrival rates $\lambda$ can reach $10^{5}$--$10^{6}$ requests
per second at peak demand~\cite{inferenceserving}.
Their challenge is not only whether capacity exists somewhere in
the cluster, but whether a scheduler can discover and commit that
capacity quickly enough before the underlying state has already
changed.

Laminar focuses on Long tasks and F-tasks after rigid-topology
jobs have already been placed.
When these tasks compete for residual capacity inside
already-fragmented nodes, macroscopic telemetry may imply
sufficient total capacity while per-node bitmap inspection
reveals no contiguous residual region of sufficient
size---a persistent \emph{false-optimism gap} that invalidates
the assumption that a scheduler can safely reason from a
quasi-static global capacity view.

This regime creates two coupled requirements.
First, Long tasks and F-tasks must be admitted with very low
hot-path control cost, because high arrival pressure and fine
time scales make heavyweight coordination self-defeating.
Second, admission correctness alone is insufficient: once these
tasks have started, they must not be abandoned to blind
host-level failure behavior whenever severe local memory
pressure arises.
In the fragmented post-landing cluster we target, scheduling
and runtime survival are therefore coupled parts of the same
ecological problem.
Within this ecology, we assume a value hierarchy: rigid-topology 
jobs are most valuable, Long tasks come next, and individual F-tasks 
are least valuable in isolation. A runtime survival policy that allows 
node-level OOM to arbitrarily kill rigid or Long jobs in order to protect 
transient F-tasks is therefore misaligned with operator intent.

\subsection{Stress Points in Representative Architectures}

Under this post-landing ecology, existing scheduling architectures
expose distinct but related stress points, because their control
paths couple resource discovery to an underlying coordination
mechanism that becomes a first-order bottleneck at the
short-task, high-arrival, high-utilization operating points we
target.

\textbf{Centralized schedulers} (e.g., Kubernetes, Slurm) pay for
global visibility through serialized coordination:
strongly consistent stores such as etcd saturate on the order of
$10^{4}$ writes per second under stable latency~\cite{etcdbenchmark}, and
global placement evaluation, even when solved by fast heuristics,
introduces solution times that grow sharply with cluster size and
constraint complexity.
Under mixed workloads at high utilization, a large job requiring
global reasoning can occupy the head of the global queue for
hundreds of milliseconds or longer, leaving F-tasks stranded with
$T_{\text{sched}} \gg T_{\text{exec}}$ despite available capacity.

\textbf{Hierarchical schedulers} (e.g., Flux) organize resources
into a recursive tree to reduce single-point pressure, but this
structure introduces an inherent coordination liability.
Sibling schedulers at the same tree level maintain isolated,
asynchronously updated state views; under high churn, their
concurrent placement decisions frequently collide at leaf nodes.
Each collision requires corrections to propagate back up the
hierarchy via RPCs, multiplying latency with each hop.
The deeper the tree and the higher the concurrency, the more
severe this rollback amplification becomes---effectively
recreating the root-level contention that partitioning was
designed to avoid.

\textbf{Local-first decentralized schedulers} (e.g., Ray) improve
median latency at moderate load, but when utilization is high,
spillback floods the Global Control Store (GCS) with RPC
redirections: in the worst case, $M$ concurrent tasks probing
across $N$ heavily loaded nodes can generate $\mathcal{O}(MN)$ control-plane
RPC attempts~\cite{ray}.
Even though each probe is small, the aggregate triggers incast at
the aggregation layer, stresses NIC receive queues, and inflates
tail latency cluster-wide.
In a cluster already fragmented by long-resident workloads, the
cost of discovering where \emph{not} to land can quickly dominate
the cost of useful work itself.

Across these architectures, the common structural problem is not
simply that they are slower in some absolute sense.
Rather, their resource-discovery logic remains tied to a
coordination mechanism whose cost grows with load, fragmentation,
or concurrency: globally visible state and locking
(I/O-bound), hierarchy-level collision correction
(structure-bound), or network-visible retry chains
(network-bound).
In the regime we target, this coupling becomes especially fragile
because Long tasks continuously reshape local ecology while
F-tasks arrive faster than heavy coordination can safely converge.

There is also a second limitation, which becomes visible once
execution begins.
In most existing paradigms, the scheduler's explicit semantic
responsibility is concentrated at admission and placement.
In the fragmented cluster ecology left behind by rigid-topology
jobs, however, post-start physical memory pressure can be just as
consequential as admission-time feasibility.
A task that was safely admitted under one local condition may
later encounter acute contention as neighboring workloads evolve.
If runtime survival is left entirely to opaque host heuristics,
declared scheduling order ceases to matter precisely when the
system is under the greatest stress.
This is not merely a runtime implementation detail; it is a
missing semantic layer in the scheduling paradigm itself.

\subsection{The Coordination Boundary and Design Requirements}

The preceding discussion suggests that next-generation AI clusters
require more than incremental improvement to an existing placement
loop.
What is needed is a scheduling paradigm whose responsibility is
defined relative to the post-landing ecology created by
rigid-topology jobs: a paradigm that keeps admission cheap for
Long tasks and F-tasks, yet does not surrender runtime survival to
undifferentiated failure behavior under severe local pressure.
Four concrete requirements follow.

\begin{itemize}

\item \textbf{Constant-cost state dissemination.}
The hot path should not depend on per-task bidirectional RPC
chains, globally serialized metadata writes, or coordination
structures whose cost scales directly with offered load.
State should instead be disseminated as lightweight aggregate
summaries so that communication cost remains effectively stable
with respect to $\lambda$ in the scale-out regime we evaluate.

\item \textbf{Local-state probing without exploratory RPC chains.}
Tasks should not have to discover capacity by injecting long
sequences of exploratory control RPCs into the physical network.
Search, congestion prediction, and collision handling should
reduce as much as possible to bounded local operations over
projected state, so that failed search remains cheap and
dissipative rather than expanding into network-wide spillback.

\item \textbf{Controlled sub-optimality under fragmentation.}
In the residual cluster left behind by rigid-topology occupancy,
globally perfect fits are often illusory or too expensive to chase.
The scheduler should therefore allow Long tasks and F-tasks to
accept sufficiently good placements into fragmented idle capacity
rather than over-optimizing against stale or overly macroscopic
state---directly addressing the false-optimism gap where aggregate
capacity signals mislead global matching under high churn.

\item \textbf{Bounded runtime survival under severe local pressure.}
Execution-start correctness is not sufficient if running tasks are
later destroyed by blind kernel-level heuristics.
The scheduling paradigm should preserve an explicit node-local
survival policy under acute physical contention: local recovery
should be preferred when the pressure spike is transient,
secondary re-addressing should remain bounded when local recovery
fails, and irreversible reclamation should occur only through a
controlled semantic path rather than arbitrary runtime collapse.

\end{itemize}

The Laminar architecture we present next is designed explicitly
around these requirements.
\section{Laminar System Design}
\label{sec:design}

\begin{figure}[htbp]
  \centering
  \includegraphics[width=\columnwidth]{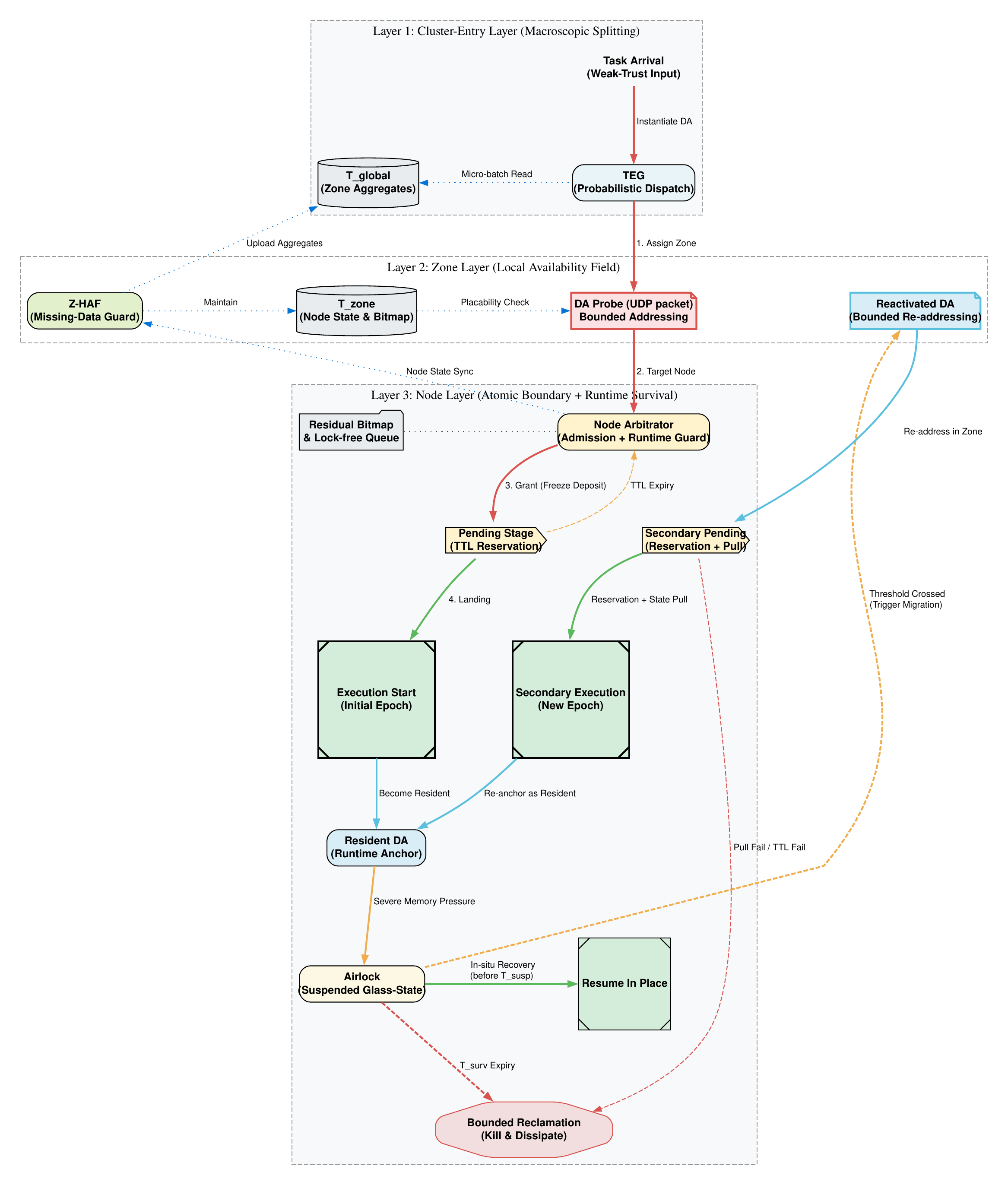}
  \caption{Overall architecture and lifecycle state machine of Laminar. The system is partitioned into three logical layers. On the downward admission path (Layers 1 and 2), TEG gateways perform macroscopic flow splitting, while DAs execute bounded in-Zone addressing utilizing Z-HAF's projected state. At the Node Layer (Layer 3), admission closes through a two-phase TTL-bounded pending stage. Crucially, the lower half details Laminar's runtime survival continuation: under severe memory pressure, the Node Arbitrator transitions Resident DAs into the \textbf{Airlock} (Suspended Glass-State). This enforces a deterministic, priority-ordered resolution routing: in-situ recovery before $T_{\text{susp}}$, threshold-triggered secondary re-addressing, or bounded reclamation upon $T_{\text{surv}}$ expiry.}
  \label{fig:laminar_arch}
\end{figure}

Figure~\ref{fig:laminar_arch} illustrates Laminar's three-layer
organization and the end-to-end control path from task arrival through
admission, execution, suspension, and, when necessary, secondary
re-addressing. The name \textit{Laminar} is intentionally
metaphorical: the goal is to make many concurrent lightweight control
objects behave in a laminar-flow-like manner at the macroscopic
level---sliding into available gaps, remaining orderly under
concurrency, and degrading by bounded dissipation rather than by retry
storms or lock contention. This effect is realized through discrete
mechanisms: probabilistic flow splitting, projected local state,
bounded node addressing, node-local arbitration, and TTL-governed
runtime recovery.

The common-case hot path remains strictly bounded through TEG,
bounded in-Zone DA addressing, node-local arbitration, and a
TTL-bounded pending stage; the cold path asynchronously propagates
Zone aggregates upward and node-state updates inward, without placing
per-task exploratory RPC chains on the critical path. Laminar operates
under \emph{weak verification}: tenant-declared priority and demand are
accepted as inputs without online truth auditing. Instead of centralized
policing, potential abuse is structurally constrained: greedy behaviors
are throttled by bounded patience budgets, bounded probe generation,
reservation expiry, and priority-ordered runtime degradation.

Critically, Laminar does not treat failed placement or runtime
displacement as faults requiring heavyweight global recovery. Failed
probes dissipate, expired reservations are reclaimed locally, and
running tasks that encounter severe physical pressure are first
suspended into a bounded Airlock state before either being resumed,
secondarily re-addressed, or irreversibly reclaimed.

\subsection{Three Layers and Four Entities}
\label{subsec:layers}

Laminar partitions the scheduling space into three logical layers with
distinct visibility and decision scope, and realizes them through four
loosely coupled entities. The \textbf{cluster-entry layer} is
intentionally restricted to Zone-level aggregate state rather than
per-node detail, so that entry-side decisions do not scale linearly
with cluster size. The \textbf{Zone layer} is the natural unit of
state propagation and mid-level coordination, within which nodes
maintain a low-cost local state field.

The \textbf{node layer} remains the atomic boundary of initial
admission: final reservation, arbitration, and execution-start
transition all close within a single node. However, Laminar does not
terminate responsibility at execution start. Once admitted, a task
remains attached to a resident DA and may later be suspended, resumed,
or secondarily re-addressed under node-local runtime control if the
host enters severe physical memory pressure.

The \textbf{Thermo-Economic Gateway (TEG)} sits at the cluster-entry
layer and performs coarse-grained probabilistic flow splitting using
Zone-level aggregates, dispatching each probe to a randomly selected
physical node (a launchpad) within the chosen Zone. The \textbf{Zone
Holographic Availability Field (Z-HAF)} sits at the Zone layer;
rather than acting as a centralized gateway, it is a replicated state
field accessible at any node, maintaining the per-node state inside a
Zone and exporting Zone-level aggregate summaries upward to TEG. The
\textbf{Decentralized Agent (DA)} is a task's lightweight control-plane
incarnation. When a tenant submits a task, the heavy payload is parked
at an upper-layer data proxy, which forwards only the scheduling
metadata (e.g., spatial demand mask and economic variables) to the
TEG. The TEG then instantiates this metadata into connectionless DA
control objects and injects them into the cluster to search for
capacity.

Upon winning local arbitration during initial admission, a node
asynchronously pulls the payload from the proxy; the proxy enforces
exactly-once semantics by serving only the first pull request and
rejecting all subsequent attempts from duplicated DAs. If execution
starts successfully, the DA does not disappear. It transitions into a
resident control object bound to the running task and remains dormant
unless runtime pressure later forces suspension and secondary
reactivation.

A task enters execution only after its DA wins node-local arbitration,
a reservation is established, payload pull succeeds within the valid
window, and execution start is observed---the concrete realization of
the probe-first, execute-later principle on the admission path. The
task lifecycle is therefore broader than initial landing alone:
arrive and instantiate as a DA; TEG assigns the DA to a physical
launchpad node within a probabilistically selected Zone; landing on
this node, the DA directly evaluates the locally replicated Z-HAF
field and performs bounded node addressing; node-local arbitration
determines whether a reservation can be granted; the task crosses into
execution; the DA becomes resident; and only later, if sustained
physical contention arises, may the same DA be reactivated for
secondary addressing.

These entities do not form a lock-based centralized pipeline. Instead,
they decompose global placement pressure into local, low-cost,
composable decisions on the common path, while preserving a bounded
runtime survival path for the exceptional case of severe node-local
memory contention.

\subsection{State Variables and the Energy Contract}
\label{subsec:variables}

Laminar uses two groups of variables coupled at runtime: nodal
thermal-state proxies and per-task economic state.

$S$ (\textbf{Slack}) denotes the current local free capacity of a
node, measured as its residual allocatable resource atoms. $H$
(\textbf{Heat}) denotes scheduling pressure as the count of pending
DAs currently queued and awaiting arbitration. At the node level,
each node periodically scans its local pending queue, counts its
in-flight DAs, and reports this count to Z-HAF as its nodal Heat.
Z-HAF aggregates per-node Heat values across all nodes in the Zone
and exports the Zone-level Heat---the total pending DA count across
the Zone---upward to TEG.

Heat therefore reflects instantaneous contention pressure at both the
node boundary and the Zone boundary: a node with many pending DAs is a
congested target that incoming probes should avoid; a Zone with high
aggregate Heat signals that the entire Zone is under heavy scheduling
load, steering TEG to route fewer new tasks there. Together, $S$ and
$H$ provide stable repulsion signals for both macroscopic routing and
microscopic addressing without requiring strongly consistent global
telemetry.

The per-task economic state is organized as follows. $M_i$ represents
the task's geometric demand as a resource-atom mask. Its scalar mass
is $m_i = \|M_i\|_1$, i.e., the number of activated resource atoms.
The initial economic quantity of task $i$ is:
\[
E_i(0) = p_i m_i, \quad m_i = \|M_i\|_1
\]
Here $p_i$ is the tenant-declared token density or priority weight
per unit mass, accepted under weak trust as a scheduling input rather
than a verified payment. At task creation, Laminar derives two runtime
quantities from the same source:
\[
E_{v,\text{init}} = E_{\text{patience}}(0) = E_i(0)
\]

$E_{v,\text{init}}$ is the static routing weight used in local
competition. $E_{\text{patience}}$ is the bounded control-plane budget
available for addressing actions, bounded retries, pending-stage
waiting, and any later secondary re-addressing if a running task is
forced into suspension. A high tenant declaration therefore
initializes both a static competition weight and a finite patience
budget, rather than granting infinitely reusable privilege.
While Laminar operates under weak online verification on the hot path, 
it inherently assumes an out-of-band upper-layer admission controller, 
quota, or billing system caps the maximum priority density ($p_i$) 
a tenant can inject over time. Laminar's responsibility is to enforce 
this declared hierarchy efficiently at runtime, not to police global 
user quotas.

The economic terminology---token density, energy, patience
budget---serves first as a control abstraction for weighting, bounded
search, and degradation ordering, while remaining compatible with
higher-layer accounting or settlement interpretations of the same
declared priority signal.

\subsection{Unified Utility Field}
\label{subsec:utility}

TEG's macroscopic flow splitting and DA's microscopic node addressing
share a common utility definition. Let $S_{\text{pred}}$ and
$H_{\text{pred}}$ denote the predicted Slack and Heat at the current
observation target, produced by Z-HAF through local time projection
as described in \S\ref{sec:zhaf}. The unified utility is:
\[
U = \log_2(1 + S_{\text{pred}}) - \gamma \log_2(1 + H_{\text{pred}})
\]
where $\gamma > 0$ controls thermal repulsion strength. Utility
increases with available capacity and decreases with pending-queue
pressure; the logarithmic form smooths extreme values and keeps the
field numerically stable.

TEG and DA differ not in the utility function they use but in how they
act on it. TEG maps Zone-level utility to a routing probability
distribution to avoid collapsing concurrent arrivals onto a single
Zone. DA performs a finite local discrete choice over sampled node
candidates to retain the tendency toward higher Slack and lower Heat
at low cost. When a suspended task later triggers secondary
reactivation, the reactivated DA re-enters the same utility field
rather than relying on a separate migration-specific addressing rule.

\subsection{TEG: Macroscopic Probabilistic Flow Splitting}
\label{subsec:teg}

When arrivals are bursty, TEG does not attempt node-level exact
matching. Instead, it performs macroscopic probabilistic flow
splitting using only Zone-level aggregate state. Let $U_z$ be the
aggregate utility of Zone $z$ and $\tau > 0$ be a temperature
parameter controlling routing sharpness. TEG assigns the dispatching
probability for Zone $z$ as:
\[
P(z) = \frac{2^{U_z / \tau}}{\sum_{r=1}^{Z} 2^{U_r / \tau}}
\]

Probabilistic splitting rather than deterministic $\arg\max$ prevents
concurrent tasks from synchronizing onto the same most attractive
Zone. At the default $\tau = 1$, the rule reduces to a ratio form in
which more Slack and less Heat increase a Zone's attraction. Missing
Zone-level state is handled conservatively: TEG does not increase the
apparent appeal of an unobservable Zone.

Because TEG reasons only over Zone-level aggregates, it remains
agnostic to whether a DA is in its initial admission epoch or in a
later secondary reactivation epoch. In both cases, TEG applies the
same macro-level routing principle: spread pressure probabilistically,
avoid synchronous herding, and leave the final decision to bounded
local addressing and node-local arbitration.

\subsection{Z-HAF: Zone State Field and Missing-Data Handling}
\label{sec:zhaf}

Z-HAF provides a local state view with temporal evolution rather than
a static snapshot. It maintains per-node state inside a Zone---
including $S$, $H$, freshness, and resource-atom availability
bitmap---and exports two Zone-level aggregates upward to TEG: the mean
residual Slack across nodes in the Zone, and the total pending DA
count across the Zone as Zone-level Heat. This ensures that TEG's
macroscopic routing signal reflects both available capacity and
current contention pressure at Zone granularity.

Let the most recently visible state of a node be $(S, H)$, let
$(\hat{\dot{S}}, \hat{\dot{H}})$ be its smoothed first-order
derivatives, and let $\tau_i$ be the sensing delay. Z-HAF forms the
projected state as:
\[
S_{\text{pred}} = \max(0,\, S + \tau_i \hat{\dot{S}}), \quad
H_{\text{pred}} = \max(0,\, H + \tau_i \hat{\dot{H}})
\]
This projection is a discrete approximation of short-term local
trends. By incorporating both current value and recent trend, Laminar
maintains a stable addressing view under asynchronous propagation,
mild delay, and short packet loss without requiring global
synchronization.

Nodes do not report state updates simultaneously. Each node fires its
report after a base interval plus a per-node Gaussian jitter
$\delta_n \sim \mathcal{N}(0, \sigma^2)$, where $\sigma$ is set to a
fraction of the base reporting interval. This jitter desynchronizes
state uploads across nodes so that Z-HAF and the upstream network do
not receive a burst of simultaneous update packets from all nodes in a
Zone, which would otherwise create incast at the aggregation layer
under high node counts.

For missing data, Z-HAF applies a \textbf{short-project, long-degrade}
rule. Brief outages are handled by projecting from the most recent
visible state and derivative. Extended silence causes Z-HAF to lower
visible Slack and raise visible Heat, conservatively removing the node
from the candidate set. This policy prioritizes avoiding false
optimism about hot or failed nodes.

The same projected-state machinery is reused when a suspended task's
resident DA is secondarily reactivated. Runtime re-addressing
therefore does not introduce a separate addressing model; it reuses
the same Zone-local predicted state semantics that already govern
Laminar's ordinary bounded search.

\subsection{DA: Persistent Lifecycle and Secondary Reactivation}
\label{subsec:da}

A DA is Laminar's basic moving unit on the control plane. However, Laminar does not treat the DA as a disposable probe whose responsibility ends once execution begins. Instead, Laminar binds one DA to one task across the task's full lifecycle. The DA may change operational mode, but it does not lose task identity: it remains the task-bound control object from initial admission, through execution, through suspension, and, if necessary, through secondary re-addressing.

\textbf{Phase I: Kinetic Addressing.}
At admission time, the DA behaves as a connectionless UDP control probe and performs a finite discrete search guided by the local utility field. Let $C$ be the sampled candidate set within the current Zone. The node-level addressing score remains
\[
\mathrm{Addr}_j = \log_2(1 + S_{\mathrm{pred},j}) - \lambda \log_2(1 + H_{\mathrm{pred},j}) + \epsilon_j,
\]
where $\epsilon_j \sim \mathcal{N}(0,\sigma^2)$ is a zero-mean perturbation used to break symmetry under concurrency. Each addressing or bounce action consumes patience. Once $E_{\text{patience}}$ can no longer support another action, the DA triggers Fast-Fail and dissipates locally.

\textbf{Phase II: Resident Sentinel.}
Once execution start is observed, the DA does not disappear. It becomes stationary rather than kinetic, serving as the task's resident control anchor at the admitted node. In this mode, the DA no longer performs active probing on the hot path, but it preserves task identity, local ownership, and the authority required for later runtime intervention. If the task completes normally, the resident DA terminates synchronously with the task.

\textbf{Phase III: Secondary Reactivation.}
If the host node later enters severe physical memory pressure and the
task is forced into suspension, the resident DA may be reactivated for
a bounded secondary addressing attempt. Upon reactivation, the DA
receives a fresh $E_{\text{patience}}$ budget and re-enters the network
as a control-plane probe, while the suspended task state remains
quiescent at the source node and both are governed by the shared
survival TTL $T_{\text{surv}}$. If this reactivated DA wins local
admission at a new node, the destination does not immediately resume
execution. Instead, it first enters the same two-phase landing
discipline used on the initial admission path: a TTL-bounded logical
reservation is established first, and the suspended execution state is
then pulled asynchronously from the source within the destination's
valid pull window. The new execution epoch is recognized only after
that pull succeeds. If no new placement is secured before
$T_{\text{surv}}$ expires, or if the reservation-to-pull transition
cannot complete within both the destination's valid window and the
remaining shared survival window, the secondary attempt fails by
bounded reclamation: both the DA and the suspended task are
irreversibly killed.

\subsection{Node Arbitration, Pending Stage, and Runtime Control}
\label{subsec:arbitration}

The Node Arbitrator converts control-plane competition into node-local
admission and runtime survival decisions. It therefore serves a dual
role: under normal conditions, it closes initial placement at a
single-node boundary; under physical stress, it becomes the task-level
runtime guardian for the workloads already resident on that node.

\textbf{Pre-Admission Physical Check.}
Before evaluating any incoming DA against the residual resource bitmap,
the Node Arbitrator actively samples the node's real-time physical
memory watermark. If physical memory remains healthy, normal admission
proceeds. However, if the sampled watermark indicates imminent physical
exhaustion or severe fragmentation pressure, the arbitrator does not
continue blindly with ordinary admission. Instead, it temporarily halts
new admission and begins a reverse-recursive degradation procedure over
the currently resident tasks.

\textbf{Admission and Pending Stage.}
Among candidates placable under the current residual resource bitmap,
the node selects the winner by comparing static routing weight
$E_{v,\text{init}}$. A task wins admission only if it is both feasible
under current residual resources and has the highest local static
priority among feasible candidates. Winning admission does not mean the
task has entered execution. A winning DA first receives a
TTL-bounded logical reservation, and a portion of
$E_{\text{patience}}$ is frozen as a deposit. This preserves Laminar's
two-stage transition: reservation first, execution recognized only
after payload pull succeeds within the valid window.

The interval between reservation grant and execution start remains the
pending stage. Waiting is not free: slow payload pull or prolonged
pre-start occupation increases the chance that the reservation expires
before execution start. If payload pull fails within the valid window,
the reservation expires, the logical hold is removed, and the frozen
deposit is forfeited.

\textbf{Reverse-Recursive Suspension.}
When the node is already under severe physical memory pressure, the
arbitrator traverses resident tasks in ascending order of static
priority $E_{v,\text{init}}$. Lower-priority tasks are suspended first
at the host level, for example through OS-supported freezer primitives
and memory-pressure relief mechanisms. Suspension proceeds only until
the node recovers above a safe memory watermark or until no lower-priority
reclaimable state remains. This reverse-recursive suspension frees
physical headroom for higher-priority tasks and prevents indiscriminate
kernel-level termination from destroying the declared priority order.

The node layer therefore remains Laminar's atomic correctness
boundary, but the boundary is now broader than admission alone. Before
granting any reservation, resuming any suspended task, or activating
any secondary re-addressing transition, the arbitrator checks a small
set of local invariants over the residual bitmap, pending state, and
suspended-state ledger. If an invariant violation is detected, the
arbitrator rejects the offending transition and triggers local recovery
rather than allowing a corrupt state transition to propagate further.

Execution start no longer marks the point at which Laminar's
responsibility ends. Instead, it marks the transition from kinetic
admission control to resident runtime governance. Under normal
execution the runtime path remains silent, but under acute memory
contention the Node Arbitrator may suspend, resume, reactivate, or
ultimately kill tasks according to Laminar's runtime survival policy.

\subsection{The Airlock Protocol: Suspension Threshold, Shared Survival TTL, and Secondary Landing}
\label{subsec:airlock}

When reverse-recursive suspension is triggered, the affected task
enters an \emph{Airlock} state rather than being immediately evicted.
The Airlock is Laminar's temporal buffer between local overload and
irreversible task death. Its purpose is to absorb short-lived physical
pressure spikes locally, while reserving heavyweight cross-node
migration for sustained starvation events.

Airlock introduces three distinct time semantics. First, a suspended
task is granted a local suspension threshold
$T_{\text{susp}}$, which defines how long the Node Arbitrator should
prefer in-situ recovery before attempting secondary re-addressing.
Second, once secondary re-addressing is activated, the suspended task
and its reactivated DA are bound to a shared survival TTL
$T_{\text{surv}}$, which is the hard upper bound on their joint
survival while the task remains suspended. Third, if the reactivated
DA later wins admission at a destination node, the subsequent state
pull is governed by the same two-phase reservation rule already used
during initial landing: the destination first obtains a TTL-bounded
logical reservation, and the suspended task state must be pulled
within that valid window before the new execution epoch is recognized.

While the task remains suspended, the Node Arbitrator continuously
monitors the node's physical recovery, producing three possible
outcomes:
\begin{enumerate}
    \item \textbf{In-situ recovery before threshold.} If memory pressure
    dissipates before $T_{\text{susp}}$ is crossed, the arbitrator
    resumes the suspended task in place. No secondary re-addressing is
    attempted, and the task returns directly to ordinary execution.

    \item \textbf{Threshold-triggered secondary reactivation.} If
    suspension persists beyond $T_{\text{susp}}$, the arbitrator
    abandons local-only recovery and reactivates the resident DA. From
    this point onward, the suspended task and the reactivated DA share
    the same survival countdown $T_{\text{surv}}$. If that countdown
    expires before a new execution epoch is successfully established,
    both the DA and the suspended task are irreversibly reclaimed.

    \item \textbf{Local death by survival expiry.} If neither local
    recovery nor successful secondary landing completes before
    $T_{\text{surv}}$ reaches zero, Laminar kills both the suspended
    task and its DA by bounded reclamation.
\end{enumerate}

State transfer is valid only while the source task remains in
suspension. If the reactivated DA wins admission at another node, the
destination does not immediately resume execution. Instead, it first
enters the same two-phase landing discipline used on the initial
admission path: reservation first, state pull second, execution epoch
recognized only after the suspended state has been pulled within the
destination's pull-valid window. This keeps runtime re-addressing
semantically aligned with Laminar's original execution boundary rather
than introducing a separate migration-specific commit rule.

The shared survival TTL $T_{\text{surv}}$ remains the source-side hard
deadline throughout this process. The destination-side pull-valid
window may govern whether the newly won landing succeeds locally, but
it may not extend the lifetime of the suspended source state beyond
$T_{\text{surv}}$. Therefore, a secondary landing succeeds only if two
conditions both hold: the destination completes its reservation-to-pull
transition within its local two-phase window, and the end-to-end
transfer finishes before the shared survival TTL expires.

This design ensures that Laminar does not react to every transient
spike with migration, but neither does it delegate survival to a blind
kernel OOM heuristic. Airlock inserts a deterministic temporal
decision layer between physical stress and task death, while preserving
the same reservation-first execution semantics that already govern
Laminar's initial landing path.

\subsection{Absolute Priority Guarantee and the Fear Premium}
\label{subsec:fear_premium}

Laminar continues to reject scalar overcommitment at admission. However, admission correctness alone is insufficient in a weak-trust environment if runtime contention is still resolved by opaque kernel heuristics. The Airlock state machine changes this by extending Laminar's declared priority order into a runtime survival order.

Under physical contention, the Node Arbitrator traverses resident tasks in ascending order of $E_{v,\text{init}}$, suspending lower-priority tasks before endangering higher-priority ones. A low-priority workload therefore faces a deterministic degradation ladder rather than arbitrary destruction: first local suspension, then a bounded window for in-situ recovery, then secondary re-addressing, and only finally irreversible reclamation if the shared survival TTL expires. High-priority workloads, in contrast, are structurally insulated from blind kernel-level termination as long as lower-priority reclaimable state still exists.

This yields an \textbf{Absolute Priority Guarantee}: declared priority is no longer merely a tie-breaker at admission time, but the exact ordering principle for survival under runtime contention. Once tenants can rely on that guarantee, the incentive to hoard excess capacity as a fear premium is reduced. The system can therefore move closer to true physical saturation without relying on protocol-level overcommitment or exposing high-priority workloads to undifferentiated OOM risk.
\section{System Implementation}
\label{sec:implementation}

This section explains how Laminar's mechanism is realized as an
entry-side control plane, a Zone-local state field, and a node-local
admission-and-runtime-survival path. Its implementation preserves the
three-layer boundary introduced in \S~\ref{sec:design}: the entry side
maintains only Zone-level aggregate state, the Zone layer maintains
node-visible state, and final admission, pending-stage closure, and
runtime survival decisions are enforced locally at a single node. The
control plane uses lightweight connectionless message passing; message
loss, reordering, and silence are absorbed by soft-state probes,
Fast-Fail, and regeneration. This section does not redefine TEG, Z-HAF,
DA, or the Node Arbitrator; it specifies the runtime tables, message
fields, and local loops that carry them.

\subsection{TEG: Entry-Side Dispatch}
\label{subsec:impl_teg}

Physically, TEG is deployed as a horizontally scalable fleet of stateless gateways behind a hardware load balancer, preventing single points of failure and ingress bottlenecks. At runtime, TEG maintains a Zone-indexed aggregate table
$T_{\text{global}}$, where each row stores the Zone's aggregate Slack,
aggregate Heat, and a freshness timestamp. These aggregates are
periodically uploaded by Z-HAF instances and refreshed at the entry
side as the direct input to flow splitting. Entry-side dispatch is not
triggered independently for each arriving task; instead, a
high-frequency send loop reads the current $T_{\text{global}}$ in
micro-batches, applies the Zone-level utility and routing rule defined in \S~\ref{sec:design}, and sends a batch of DAs toward their selected Zones.
TEG therefore operates as two coupled loops: a lower-frequency refresh
loop for Zone summaries and freshness, and a higher-frequency
micro-batch dispatch loop for stable ingress throughput.

To deliver the first control packet, TEG also maintains a lightweight
node endpoint table that records a node identifier, address, and health
bit for each candidate first-hop node in each Zone. This table solves
only the first-hop delivery problem; node-level feasibility remains
determined later by $T_{\text{zone}}$ and node-local state.

\subsection{Z-HAF: Zone-State Aggregation}
\label{subsec:impl_zhaf}

In our deployment topology, a Zone maps to a bounded physical fault domain (typically an aggregation pod or a leaf-spine block). Within this domain, Z-HAF is deployed as a lightweight state aggregator-reflector service. To support fully decentralized addressing without a centralized gateway, the node-state matrix $T_{\text{zone}}$ is replicated locally in the physical memory of every node within the Zone. Its rows store the node's current $S$, $H$, freshness, and resource-atom availability bitmap, optionally accompanied by smoothed first-order derivatives for short-term prediction. A resource atom is the
predefined minimum allocatable unit inside a node, such as a CPU core,
GPU slot, NUMA region, or bandwidth bucket. Because the task mask and
the node's atom-availability bitmap live in the same semantic space, DA
can directly use $T_{\text{zone}}$ to test local placability; in
implementation terms, $T_{\text{zone}}$ is both a Zone snapshot and
the direct lookup structure for bounded addressing.

Each node computes its Heat as the current count of pending DAs in its
local arbitration queue. Nodes do not report every local state change
immediately; instead, each node aggregates local changes within a short
window and uploads them to Z-HAF in micro-batches. To prevent
synchronized bursts of state reports from all nodes in a Zone arriving
simultaneously at Z-HAF---which would otherwise create incast at the
aggregation layer under high node counts---each node schedules its
upload after the base reporting interval plus a per-node Gaussian jitter
$\delta_n \sim \mathcal{N}(0, \sigma^2)$, where $\sigma$ is a
configurable fraction of the base interval. Z-HAF uses these staggered
updates to refresh $T_{\text{zone}}$ and then produces Zone-level
aggregate Slack (mean residual across nodes), Zone-level aggregate Heat
(total pending DA count across the Zone), and freshness for TEG.

Z-HAF therefore serves a dual role in the runtime path: upward, it
exports macroscopic summaries to the entry side; downward, it exposes
the node-visible state used by DA for projection and addressing. Under
short-lived staleness, DA combines the last visible state, derivative
terms, and sensing delay to form $S_{\text{pred}}$ and $H_{\text{pred}}$,
reducing the effect of asynchronous propagation. Under prolonged
silence, Z-HAF conservatively lowers visible Slack or raises visible
Heat so that unobservable nodes do not become artificially attractive.
This downward exposure is achieved by Z-HAF periodically broadcasting the complete refreshed $T_{\text{zone}}$ back to all physical nodes. Zone-internal dissemination uses lightweight UDP messages; when hardware multicast is available it may assist fan-out, otherwise the same interface falls back to multi-unicast or lightweight gossip.

\subsection{DA: Persistent Lifecycle and Secondary Reactivation}
\label{subsec:impl_da}

A task is instantiated at the admission boundary as a DA and encoded as
a fixed-length lightweight control message carrying the task ID,
$E_{v,\text{init}}$, $E_{\text{patience}}$,
$\mathrm{TTL}^{\text{probe}}$, and a
resource-atom-aligned demand mask. The mask represents geometric
occupancy rather than scalar size, and each bit corresponds to a
predefined resource atom; the task mass $m_i$ is therefore obtained
directly from the number of activated bits. The tenant-declared token
weight that contributes to $E_i(0)$ is not strongly verified online;
it is accepted as weak-trust scheduling input, and its effect is
constrained through bounded search, reservation, pending-stage cost,
and conservative visibility rules.

After creation, the DA is first handed to TEG, which selects a target Zone using the current aggregate state and sends the initial control packet to a concrete first-hop physical node (launchpad) through the endpoint table. Landing on this launchpad node, the DA directly reads the local in-memory $T_{\text{zone}}$ replica to perform local placability checks, bounded candidate sampling, and computes the optimal target $j^\star$. If the launchpad itself is suboptimal, the DA performs a single-hop physical redirection (bounce) to $j^\star$. Each addressing or physical bounce action incurs a cost; if the remaining patience budget can no longer support the next action, the DA triggers Fast-Fail and is reclaimed locally, so the search process remains finite rather than turning into an unbounded retry chain.

Once execution start is observed, the DA does not disappear. Instead,
it transitions from a kinetic probe into a resident task-bound control
object anchored at the admitted node. In this resident mode, the DA no
longer performs active probing on the hot path, but it preserves task
identity, local ownership, and the authority required for later runtime
intervention.

The resident DA is maintained in a compact local runtime table keyed by
task identifier and execution epoch. Its state records at least the
task ID, current mode, owning node, source execution epoch, and the
control metadata required for possible secondary reactivation. If the
task completes normally, the resident DA is retired synchronously with
the task.

If the host later enters severe physical memory pressure and the task is
forced into suspension, the resident DA may be reactivated for a bounded
secondary addressing attempt. Upon reactivation, the DA receives a fresh
$E_{\text{patience}}$ budget and re-enters the network as a control-plane
probe, while the suspended task state remains quiescent at the source
node and both are governed by the shared survival TTL
$T_{\text{surv}}$.

If the reactivated DA wins local admission at a new node, the
destination does not immediately resume execution. Instead, it first
enters the same two-phase landing discipline used on the initial
admission path: a TTL-bounded logical reservation is established first,
and the suspended execution state is then pulled asynchronously from the
source within the destination's valid pull window. The new execution
epoch is recognized only after that pull succeeds.

If no new placement is secured before $T_{\text{surv}}$ expires, or if
the reservation-to-pull transition cannot complete within both the
destination's valid window and the remaining shared survival window, the
secondary attempt fails by bounded reclamation: both the DA and the
suspended task are irreversibly killed.

Because DA remains soft state, duplicate secondary attempts may still
temporarily exist under regeneration or delayed observation. Laminar
therefore advances only the first DA that completes the destination-side
reservation-to-pull transition successfully; all late or redundant
winners dissipate by timeout or are reclaimed when the shared survival
window expires. Zone-locality is therefore not a property of the task's
entire lifetime, but of each execution epoch recognized by Laminar.

\subsection{Node: Local Arbitration, Airlock, and Runtime Control}
\label{subsec:impl_node}

Physically, a Node in Laminar corresponds to a bare-metal worker server,
and the Node Arbitrator runs as a local background daemon (or user-space
process) on the host OS. At the node boundary, incoming DAs are
intercepted by a lightweight fast path and enqueued into a local
lock-free queue for batch arbitration. The node maintains a residual
resource bitmap representing the currently allocatable resource atoms
and serving as the direct feasibility view for reservation decisions.
The current queue depth is also the node's Heat value reported
periodically to Z-HAF, as described above.

Before each admission round, the Node Arbitrator also samples a local
physical memory watermark and a compact fragmentation-pressure signal
derived from the host runtime. If the node remains within a safe
operating envelope, ordinary admission proceeds unchanged. If the node
enters severe physical memory pressure, however, the arbitrator does not
continue blind admission. Instead, it temporarily throttles new
admissions and enters a reverse-recursive runtime control path over the
currently resident tasks.

Under ordinary admission, the node considers candidates that are
placable under the current residual bitmap and selects the winner by
comparing static routing weights $E_{v,\text{init}}$. Once a reservation
is granted, the corresponding atoms are removed from the residual bitmap
before the node proceeds to the next feasible candidate. Node-local
ordering is therefore jointly determined by placability under current
residual resources and static routing weight, and the entire admission
path remains local to a single node.

The interval between reservation grant and execution start is the
pending stage. During this stage, the task occupies node-local state
without yet becoming a running task, and its frozen deposit remains
locked. Waiting is therefore not free: slow payload pull, delayed
landing, or prolonged pre-start occupancy accumulates pending-stage cost
before the hard timeout is reached. The destination-side pull-valid
window remains class-dependent and is chosen to cover high-percentile
landing latency for the relevant task class. When a reservation is
created, the node registers a timeout entry in a local timing wheel for
expiration at $t_{\text{now}} + \mathrm{TTL}^{\text{pull}}_i$; if
execution start is observed within the valid window, that entry is
cancelled. Otherwise, expiration triggers local recovery: the
reservation is removed, the residual bitmap is restored, waiting context
is cleaned up, and the frozen deposit is forfeited.

When severe memory pressure persists at runtime, resident tasks are
traversed in ascending order of static routing weight
$E_{v,\text{init}}$. Lower-priority tasks are suspended first and
entered into a local Airlock ledger rather than being immediately
killed. Each Airlock entry records at least the task identity, source
execution epoch, suspension timestamp, suspension threshold
$T_{\text{susp}}$, remaining shared survival TTL $T_{\text{surv}}$, and
the metadata required to reactivate the resident DA if local recovery
does not occur in time.

While a task remains in Airlock, the node monitors two concurrent local
timers. If physical pressure dissipates before $T_{\text{susp}}$ is
crossed, the task is resumed in place and its resident DA returns to
stationary mode. If suspension persists beyond $T_{\text{susp}}$, the
resident DA is reactivated for bounded secondary addressing while the
suspended task state remains quiescent at the source node. If
$T_{\text{surv}}$ expires before either in-situ recovery or successful
secondary landing completes, both the DA and the suspended task are
reclaimed by bounded local cleanup.

In a Linux-based realization, Airlock entry can be implemented by
placing the selected low-priority task into a cgroup v2 freezer state
and invoking compressed memory relief through zRAM-backed reclaim. This
produces a compact quiescent suspended image at the source node while
preserving task identity and restart authority for the resident DA. We
informally refer to this host-level suspended image as a glass-state.
However, Laminar's mechanism does not depend on this specific kernel
path: any equivalent substrate that can quiesce the task and retain a
bounded recoverable suspended state is sufficient.

If a reactivated DA later wins local admission at a destination node,
the destination does not immediately resume execution. Instead, it first
obtains a TTL-bounded logical reservation and then pulls the suspended
state from the source within the destination's valid pull window. The
new execution epoch is recognized only after that pull succeeds.
Therefore, the source-side shared survival TTL and the destination-side
reservation-to-pull window jointly define secondary landing success: the
transfer must complete within both bounds.

The node layer remains Laminar's atomic correctness boundary. Before
granting any reservation, suspending any resident task, resuming any
Airlock entry, or activating any secondary re-addressing transition, the
Node Arbitrator checks a small set of local invariants over the residual
bitmap, pending state, resident-DA table, and suspended-state ledger.
If an invariant violation is detected, the arbitrator rejects the
offending transition and triggers local recovery rather than allowing a
corrupt reservation or runtime state transition to proceed. All
expected failure modes---false optimism about placability, patience
exhaustion, pending-stage timeout, suspension-threshold crossing,
survival-window expiry, and bounded regeneration limits---are handled as
soft-state events that trigger local reclamation without breaking the
arbitration loop.

\subsection{Runtime Closure}
\label{subsec:impl_closure}

The minimum engineering closure of the extended Laminar runtime consists
of nine components: $T_{\text{global}}$, the node endpoint table,
$T_{\text{zone}}$, fixed-length DA control messages carried over UDP,
the residual bitmap with lock-free local queue, the resident-DA runtime
table, the local Airlock/suspended-state ledger, the host
memory-pressure sampler, and the timing-wheel subsystem that enforces
both reservation expiry and runtime survival deadlines. On top of this
minimal path, the system may add optional optimizations such as kernel
bypass, vectorized bitmap operations, or hardware multicast; these
reduce local control-path cost but do not change the mechanism itself.
Optional safeguards such as probe admission rate limiting and
conservative missing-data handling enlarge the safe operating region
under flooding or memory pressure but are not preconditions for the
Laminar mechanism to exist.

Within this closure, the runtime path is consistent end-to-end: a task
becomes a DA at admission; TEG micro-batches it using
$T_{\text{global}}$; Z-HAF maintains $T_{\text{zone}}$ through
staggered micro-batched node updates carried over UDP; the DA performs
bounded addressing inside the selected Zone; the node grants a
TTL-bounded reservation and enforces pending-stage recovery via timing
wheel; and, after execution start, the resident DA normally remains
silent unless node-local memory pressure triggers Airlock entry,
threshold-based secondary reactivation, in-situ resume, or bounded
reclamation. This preserves the probe-first, execute-later landing
boundary while extending Laminar with a local runtime survival path
under acute physical stress.

\subsection{Architectural Pathways for Hardware Offloading}
\label{subsec:hardware_offload}

The current implementation of Laminar (\S\ref{sec:implementation})
utilizes standard Linux networking and host-runtime facilities to
validate the decentralized scheduling paradigm. Because Laminar's
control plane fundamentally decouples resource discovery from complex
state-machine synchronization---relying instead on stateless probes,
bounded local scoring, and node-local closure---the architecture remains
conceptually compatible with data-plane network hardware. This section
outlines several architectural pathways for accelerating Laminar's
macroscopic and microscopic components in hardware without changing the
mechanism itself.

\textbf{Zone-Level Fanout via Hardware Multicast.}
In the software realization, TEG delivers the initial control packet
through a gateway-maintained endpoint table, while Zone-internal
dissemination uses lightweight UDP-based mechanisms. Hardware multicast
can therefore be viewed as an acceleration path for selected fan-out
operations rather than as a semantic dependency of Laminar's control
plane. In a suitable fabric, multicast-assisted dissemination could
reduce host-side replication cost for Zone-local state exposure and
selected probe fan-out patterns.

\textbf{In-Network TEG via Programmable Switches.}
The Thermo-Economic Gateway (TEG) avoids deterministic task-to-node
matching and instead performs statistical routing from Zone-level
aggregate state. This macroscopic logic aligns naturally with
match-action pipelines in programmable switches. In principle, a
quantized form of the Zone utility field could be maintained in switch
SRAM, while ingress routing could be implemented through weighted
multipath selection. Such an offload would move part of entry-side flow
splitting into the network data plane while preserving Laminar's
Zone-granular visibility boundary.

\textbf{Node Arbitration via Kernel Bypass or SmartNIC Assist.}
At the node boundary, the Node Arbitrator evaluates fixed-length demand
masks using bounded bitmap operations and short local queues. This
computation path is structurally compatible with kernel-bypass
mechanisms, SmartNIC steering, or cache-friendly user-space receive
paths. These optimizations can reduce local arbitration latency, but
they do not alter Laminar's semantics: placement still closes at a
single node, pending-stage recovery remains TTL-bounded, and runtime
survival decisions remain source-local until a secondary landing is
successfully completed.

\textbf{Runtime Survival Assist.}
The extended runtime path also exposes a second hardware acceleration
surface: host-assisted detection and enforcement of memory-pressure
events. For example, memory watermark sampling, timing-wheel expiry, and
resident-DA wakeup paths could be partially accelerated through
specialized host controllers or SmartNIC coordination. Even in such a
realization, however, Airlock remains a local runtime control policy
rather than an in-network migration protocol: suspension, shared
survival accounting, and destination-side reservation-to-pull closure
must still obey the same semantics defined by Laminar's software
mechanism.

\section{Evaluation}

We organize the evaluation around five experiments that follow an
outside-in structure. We first compare Laminar against Slurm-like,
Flux-like, and Ray-like scheduling under increasing offered load to
establish where existing designs become coordination-bound,
structure-bound, or retry-bound (Exp1). We then examine Laminar alone
across cluster scales from 5{,}000 to 32{,}000 nodes to test whether
its hot-path control cost remains approximately constant as the cluster
grows (Exp2). Next we stress Laminar's own state visibility by sweeping
synchronization delay to assess how much staleness the probe-first,
late-binding architecture can absorb before degrading (Exp3). We then
dissect two specific mechanisms---two-phase reservation and DA
regeneration---to isolate their individual contributions to Laminar's
robustness under pending-state interference and probe loss (Exp4).
Finally, we evaluate Laminar's Airlock runtime-survival layer under
sustained node-local memory pressure to test whether it converts opaque
post-start OOM destruction into bounded dissipation while preserving a
high execution-survival ratio (Exp5). All five experiments use the
setup described in \S\ref{sec:setup}.

\subsection{Experimental Setup}
\label{sec:setup}

\textbf{Simulation environment and calibration.}
All experiments use a discrete-event simulator whose timing
parameters are derived from micro-benchmarks on a standard compute
instance (2~vCPU, Intel Xeon Platinum 8369B, 2.70~GHz base),
each measured over 10 independent runs with the p50 median. The
environment enforces a 0.5~ms network RTT and 1\% physical
control-packet loss. Core hot-path costs: AVX2 bitmap feasibility
check 4.02~ns; DA utility scoring 13.7~ns; Zone-level aggregation
29.3~ns. Infinite queuing is disabled in all experiments so that
saturation manifests as observable latency and success-ratio
degradation.

\textbf{Workload.}
All experiments inject tasks as a pure open-loop Poisson arrival
stream at rate $\lambda$, independent of cluster response time.
In a closed-loop model, task generation is implicitly throttled
by system latency, suppressing the backpressure that would
otherwise expose control-plane saturation. The open-loop design
removes this masking: as offered load $\rho = \lambda / \mu$
approaches saturation (where $\mu$ is the cluster's maximum
sustainable throughput under ideal conditions), any scheduler
whose hot path is coupled to global locking, hierarchical
synchronization, or unbounded retry chains is forced to confront
non-linear queue growth directly. $\rho = 1.0$ corresponds to
full saturation.

The workload is bimodal, composed of two task classes that
jointly stress both the control plane and the node-local
resource layout.

\textit{Fine-grained transient tasks} ($\mathcal{F}$-tasks,
80\% of arrivals) have exponentially distributed service times
with a mean in the low-millisecond range and require a small
number of resource atoms that may be dispersed within a node's
bitmap. Their primary stress role is to generate a sustained
high-frequency signaling load and, through rapid small
allocations, progressively fragment the contiguous free regions
within each node's bitmap.

\textit{Large-footprint tasks} ($\mathcal{L}$-tasks, 20\% of
arrivals) have lognormally distributed service times with a
heavier tail and require a large, strictly contiguous block of
resource atoms within a single node's bitmap. 

Each node's resources are represented as a fixed-length binary
bitmap; all feasibility checks and allocations are resolved
through bitwise operations, natively embedding spatial
fragmentation into the scheduling path. This abstraction
reproduces the false-optimism gap identified in
\S\ref{sec:motivation}: a node may appear to have sufficient
aggregate capacity in a coarse summary while its bitmap
contains no contiguous free region large enough to satisfy an
incoming $\mathcal{L}$-task demand mask, forcing a probe
bounce that would be invisible to any scheduler reasoning from
scalar capacity signals alone.

\textbf{Baseline modeling.}
Rather than running specific open-source codebases, we implement
three scheduling paradigms as optimistic physical models, which
incorporate optimistic locality and multi-entry ingress, to isolate
architectural bottlenecks from engineering artifacts. All three
models are deliberately optimistic \emph{in favor of the
baseline}: engineering-level inefficiencies such as etcd WAL
fsync latency, TCP retransmission overhead, and GCS serialization
cost are omitted, and per-operation timing is set to idealized
values well below measured hardware limits. Any performance gap
favoring Laminar under these conditions is therefore a lower bound
on the gap under realistic deployment. All three baselines share
the same network ground rules as Laminar: inter-node hop delay
0.5~ms and a 10~ms global heartbeat for state synchronization.

\textit{Slurm-like} instantiates the coordination-bound
bottleneck identified in \S\ref{sec:motivation}: a globally
serialized scheduler that maintains a single authoritative
resource view and a strict global queue. Per-node resource scan
cost is set to 0.01~$\mu$s and base match cost to 0.1~$\mu$s,
giving the model every computational advantage. Despite this,
the model enforces the architecture's unavoidable physical
constraint: every placement decision must hold a global mutex
(base cost 0.5~$\mu$s), and when concurrent queue depth exceeds
10{,}000, a non-linear lock-convoy penalty activates. Tasks that
lose contention retry up to 3~times at 2~ms base backoff. Unlike
Ray-like and Flux-like, which use active messaging and incur
real resource costs from long-pending network state, Slurm-like
is a passive global queue that does not generate active signals
while waiting; the 5{,}000~ms task timeout is therefore not
applied, allowing unbounded in-memory queuing as a deliberate
concession in favor of the baseline. As shown in
\S\ref{sec:exp1}, this concession does not help: the global
lock bottleneck drives p99 latency to the simulation
horizon regardless.

\textit{Ray-like} instantiates the retry-bound bottleneck: a
local-first design with global spillback via a sharded GCS. RPC
serialization, Actor lifecycle overhead, and GCS transaction
latency are fully removed; local processing costs 20~$\mu$s and
GCS processing 50~$\mu$s. Three structurally inescapable
constraints are preserved. (1)~\textit{Local mutual exclusion}:
reservations serialize through a per-node commit lock.
(2)~\textit{State staleness and spillback}: the GCS view
synchronizes only via the 10~ms heartbeat; capacity exhaustion
forces spillback at 0.5~ms per redirect. (3)~\textit{USL
contention}: the GCS is partitioned into 32~shards with 0.5
hotspot skew; when spillback load exceeds 500~queued tasks, a
Universal Scalability Law penalty activates, reproducing the
superlinear throughput collapse from coherence overhead---the
system-level consequence of the $\mathcal{O}(MN)$ RPC
amplification described in \S\ref{sec:motivation}.

\textit{Flux-like} instantiates the structure-bound bottleneck:
a hierarchical broker tree with fanout~16 and leaf capacity~32~\cite{flux}.
Multi-level graph-matching is fully removed; parent--child
dispatch costs 1~$\mu$s and leaf scan 0.005~$\mu$s. Three
topology-level laws are enforced. (1)~\textit{Root choke point}:
all dispatches and retries pass through high-level brokers; above
4{,}000 concurrent tasks an exponential congestion penalty
activates at the root. (2)~\textit{Isolated ledgers}: sibling
brokers share no memory and decide from stale views updated only
via the 10~ms heartbeat. (3)~\textit{Cascading rollback}:
leaf-node collisions cannot resolve laterally; displaced tasks
traverse back toward the root at 0.5~ms per hop plus 10~ms base
backoff per level---the rollback amplification path described in
\S\ref{sec:motivation}.

\textbf{Laminar parameterization.}
Laminar's control-plane parameters are set as follows. Z-HAF sensing delay $\tau_i = 10.0$~ms, matching the heartbeat
granted to all baselines. DA probes carry a 150~ms silence TTL;
candidate evaluation costs 3.0 patience units, a bounce costs
6.0~units, and Fast-Fail triggers below a 1.0-unit floor.
Inter-regeneration quiet interval is 150~ms with a 5-instance
cap. Winning probes enter the pending stage with a 50-unit frozen
deposit and a 200~ms payload-pull TTL. The absolute task timeout
is 500~ms---one tenth of the 5{,}000~ms granted to the three
baselines, which require multi-level queuing and rollback chains
to resolve contention. The baselines therefore receive both
optimistic architectural treatment and a generous survival window;
Laminar receives neither. Any observed advantage is a lower bound
on the gap under realistic deployment.

\subsection{Mixed-Load Comparison (Exp1)}
\label{sec:exp1}

We run all four paradigms on a 5{,}000-node heterogeneous cluster
with 256-node Zones and 20\% intra-Zone size jitter under a
bimodal open-loop workload, sweeping offered load $\rho$ from 0.4
to 0.9 over 30~s simulated time. Laminar runs with Z-HAF
Taylor projection, missingness guard, and DA regeneration enabled;
two-phase reservation is disabled to isolate hot-path behavior from
pending-state recovery. All baselines share the same network ground
rules as described in \S\ref{sec:setup}.

\begin{figure}[htbp]
  \centering
  \begin{minipage}[t]{0.48\columnwidth}
    \centering
    \includegraphics[width=\linewidth]{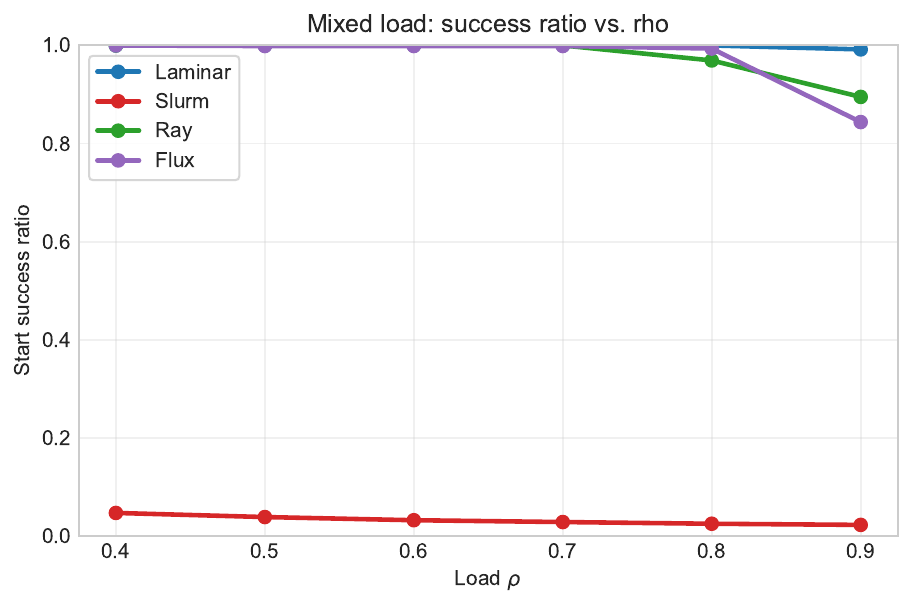}
  \end{minipage}
  \hfill
  \begin{minipage}[t]{0.48\columnwidth}
    \centering
    \includegraphics[width=\linewidth]{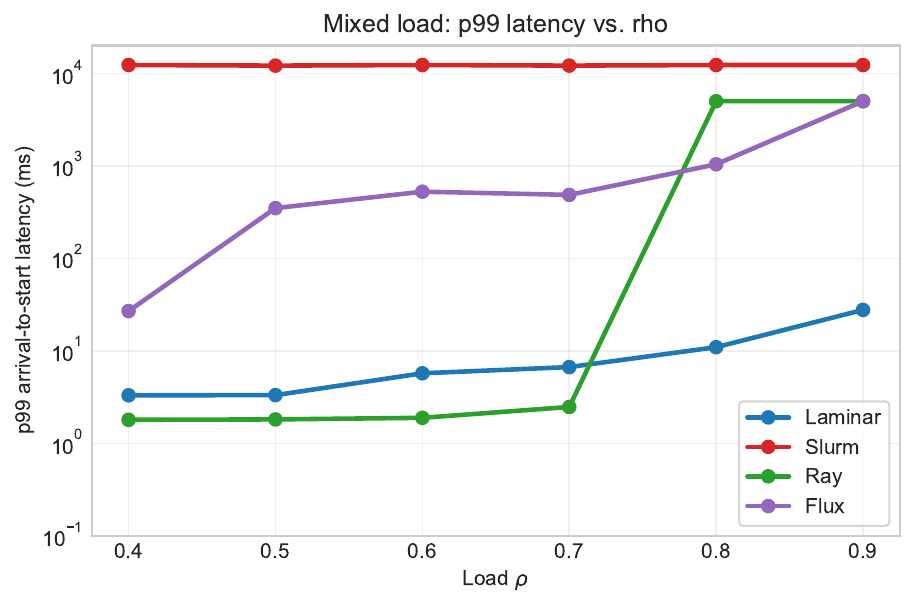}
  \end{minipage}
  \caption{Start success ratio (left) and p99 arrival-to-start
  latency (right) for four scheduling paradigms under mixed load,
  $\rho \in \{0.4, 0.5, 0.6, 0.7, 0.8, 0.9\}$, 5{,}000-node cluster.}
  \label{fig:exp1}
\end{figure}

Figure~\ref{fig:exp1} compares Laminar, Slurm-like, Flux-like, and
Ray-like scheduling under mixed load as $\rho$ increases from 0.4
to 0.9. Slurm-like scheduling is already saturated across the
entire sweep: its p99 arrival-to-start latency stays around
12.1--12.3~s while start success ratio remains only
2.3\%--4.8\%, confirming that a globally serialized decision
path is incompatible with the high-arrival, short-task regime we
target. Flux-like and Ray-like both perform well at low load but
degrade sharply as $\rho$ rises: Flux-like p99 grows from
27.05~ms at $\rho=0.4$ to 1{,}041~ms at $\rho=0.8$, then reaches
the 5{,}000~ms cap at $\rho=0.9$ with success ratio collapsing to
84.4\%; Ray-like stays near 1.81--2.49~ms through $\rho=0.7$ but
also hits the cap at $\rho=0.8$ with success ratio at 96.9\%,
consistent with the spillback amplification described in
\S\ref{sec:motivation}. Laminar's p99 increases from 3.33~ms at
$\rho=0.4$ to 11.01~ms at $\rho=0.8$ and 27.84~ms at $\rho=0.9$,
while start success ratio stays at 99.99\% through $\rho=0.8$ and
99.18\% at $\rho=0.9$. This gradual degradation reflects the
bounded nature of Laminar's hot path: entry-side routing depends
only on Zone summaries, in-Zone search is patience-bounded, and
placement contention closes at the node boundary rather than
propagating outward as retries or rollbacks.

\subsection{Scale-Out Behavior (Exp2)}
\label{sec:exp2}

Having established where competing designs saturate, we test
whether Laminar's hot-path cost remains approximately constant as
the cluster itself grows. We run Laminar alone at fixed $\rho=0.8$
across four cluster sizes---5{,}000, 10{,}000, 20{,}000, and
32{,}000 nodes over 30~s---with Zone size fixed at 256 nodes so that Zone
count scales proportionally with cluster size. All other parameters
are identical to Exp1.
\begin{figure}[htbp]
  \centering
  \begin{minipage}[t]{0.48\columnwidth}
    \centering
    \includegraphics[width=\linewidth]{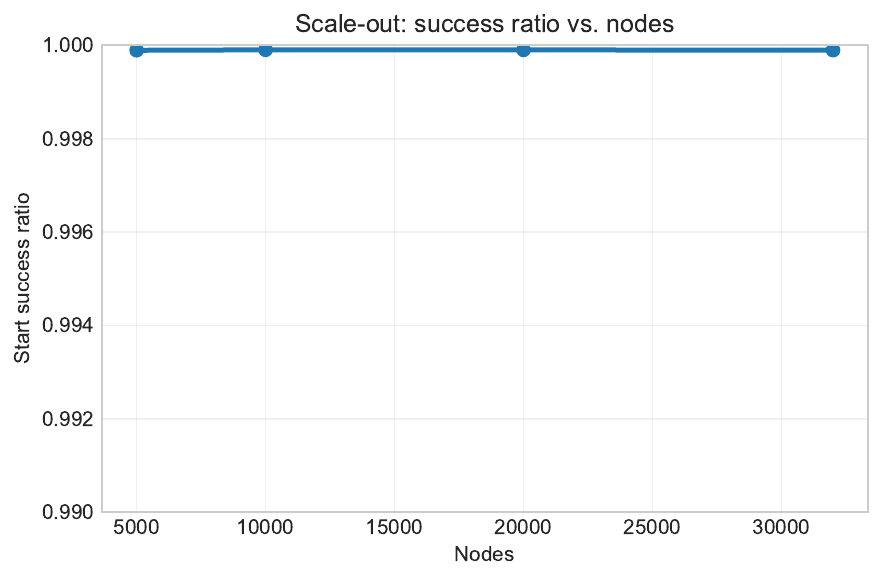}
  \end{minipage}
  \hfill
  \begin{minipage}[t]{0.48\columnwidth}
    \centering
    \includegraphics[width=\linewidth]{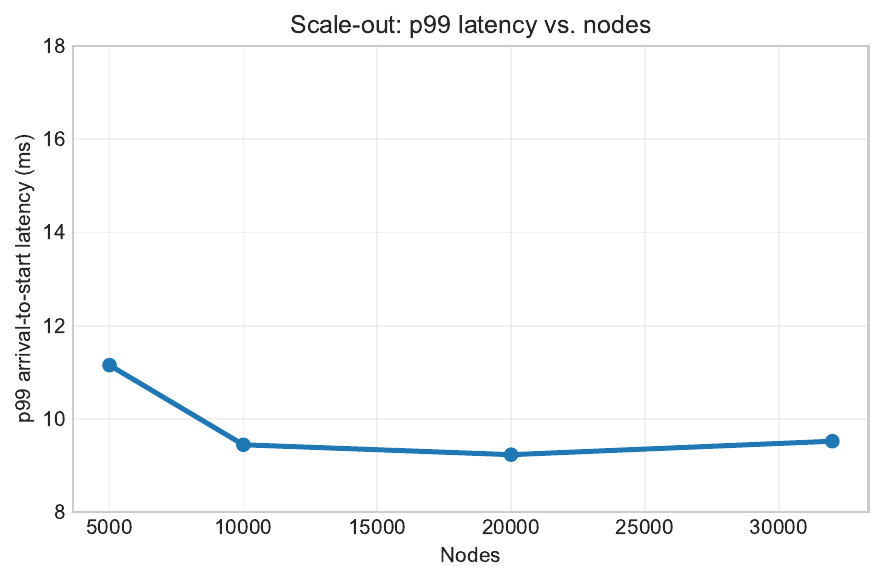}
  \end{minipage}
  \caption{Start success ratio (left) and p99 arrival-to-start
  latency (right) for Laminar as cluster size scales from
  5{,}000 to 32{,}000 nodes at fixed $\rho=0.8$.}
  \label{fig:exp2}
\end{figure}
Figure~\ref{fig:exp2} shows that p99 latency remains within
9.24--11.16~ms and success ratio stays at 99.99\% across all four
scale points. Notably, latency does not increase with node count:
p99 at 20{,}000 nodes (9.24~ms) is marginally lower than at
5{,}000 nodes (11.16~ms), reflecting natural load spreading across
more Zones rather than any lengthening of the hot path. This is
the expected consequence of the three-layer structure: TEG reads
only per-Zone aggregate summaries regardless of cluster size, DA
performs a bounded candidate scan within a single Zone, and
node-local arbitration closes placement without any cross-Zone
coordination. Increasing the total node count therefore does not
lengthen the hot path, consistent with the approximately
$\mathcal{O}(1)$ control-work claim stated in \S\ref{sec:intro}.

\subsection{Control-Plane Cost Stability}

\begin{figure}[htbp]
  \centering
  \begin{minipage}[t]{0.48\columnwidth}
    \centering
    \includegraphics[width=\linewidth]{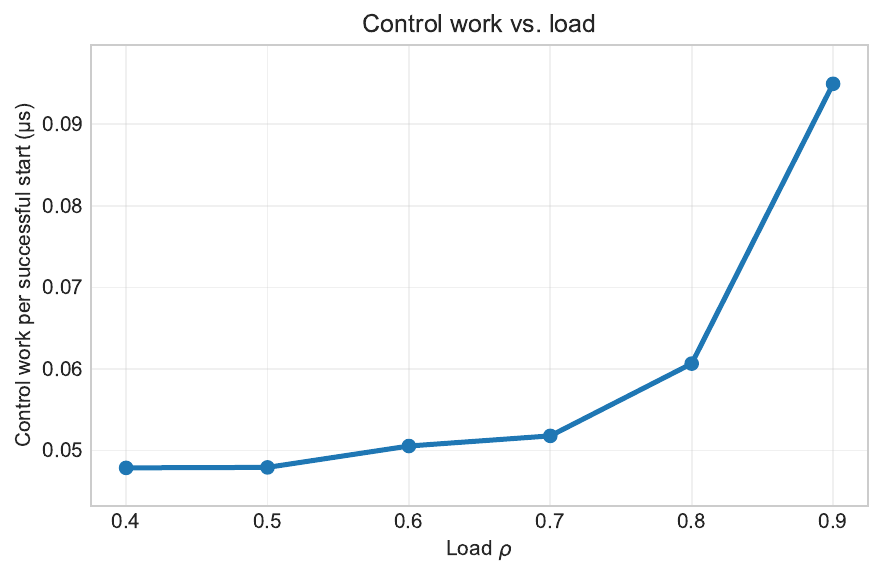}
  \end{minipage}
  \hfill
  \begin{minipage}[t]{0.48\columnwidth}
    \centering
    \includegraphics[width=\linewidth]{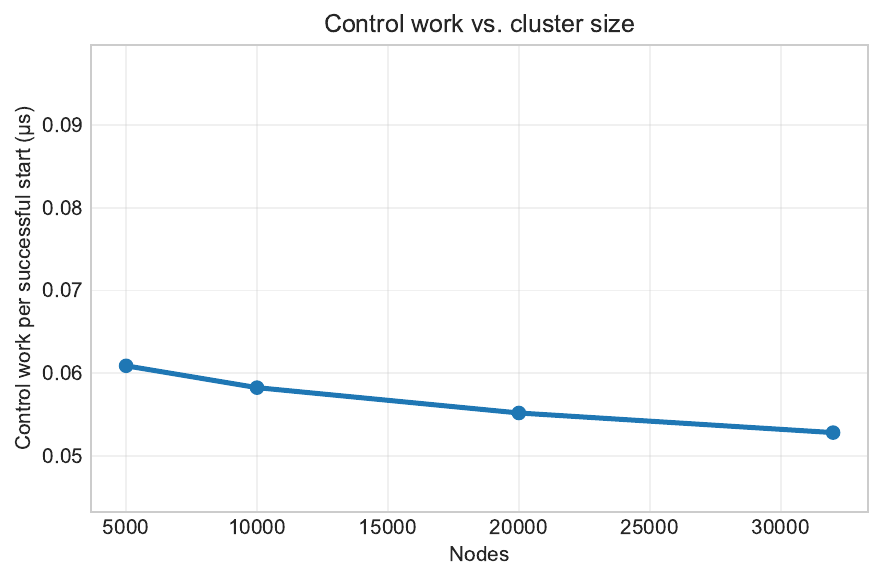}
  \end{minipage}
  \caption{Control work per successful execution start for Laminar under mixed load (left) and under scale-out at fixed $\rho=0.8$ (right). In the mixed-load sweep, the measured control work rises only mildly from $0.0479\,\mu s$ at $\rho=0.4$ to $0.0950\,\mu s$ at $\rho=0.9$. In the scale-out sweep, it remains tightly bounded, decreasing from $0.0609\,\mu s$ at 5{,}000 nodes to $0.0528\,\mu s$ at 32{,}000 nodes.}
  \label{fig:control_work}
\end{figure}

To make the near-\(O(1)\) hot-path claim more explicit, we additionally report control work per successful execution start across both the mixed-load sweep and the scale-out sweep. As shown in Fig.~\ref{fig:control_work}, Laminar's control work grows only mildly as offered load increases. Across the main operating range, the measured value rises from \(0.0479\,\mu s\) at \(\rho=0.4\) to \(0.0950\,\mu s\) at \(\rho=0.9\), with intermediate points remaining similarly small, indicating that even under higher utilization the common path does not experience a coordination blow-up.
A similar pattern appears in the scale-out sweep. At fixed \(\rho=0.8\), the control work per successful execution start remains tightly bounded as the cluster grows, decreasing from \(0.0609\,\mu s\) at 5{,}000 nodes to \(0.0528\,\mu s\) at 32{,}000 nodes. This result complements the success-ratio and latency curves by showing that Laminar's stability is not obtained by hiding increasing coordination effort behind similar end-to-end outcomes; rather, the underlying control-plane work itself remains approximately constant in the evaluated scale-out regime.

This behavior is consistent with Laminar's architectural decomposition: entry-side routing depends only on Zone-level aggregates, DA search remains patience-bounded within a Zone, and placement closes at a single node boundary.

\subsection{State Staleness Tolerance (Exp3)}
\label{sec:exp3}

Scale-out stability assumes that Zone-level state remains
reasonably fresh. To test how much synchronization delay Laminar
can absorb before its behavior degrades, we fix the cluster at
5{,}000 nodes and $\rho=0.8$, and sweep
Z-HAF synchronization delay across six settings: 0, 5, 10, 20,
50, and 100~ms over 30~s. The delay is injected directly into the Z-HAF
state update path; all other parameters match Exp1.

\begin{figure}[htbp]
  \centering
  \begin{minipage}[t]{0.48\columnwidth}
    \centering
    \includegraphics[width=\linewidth]{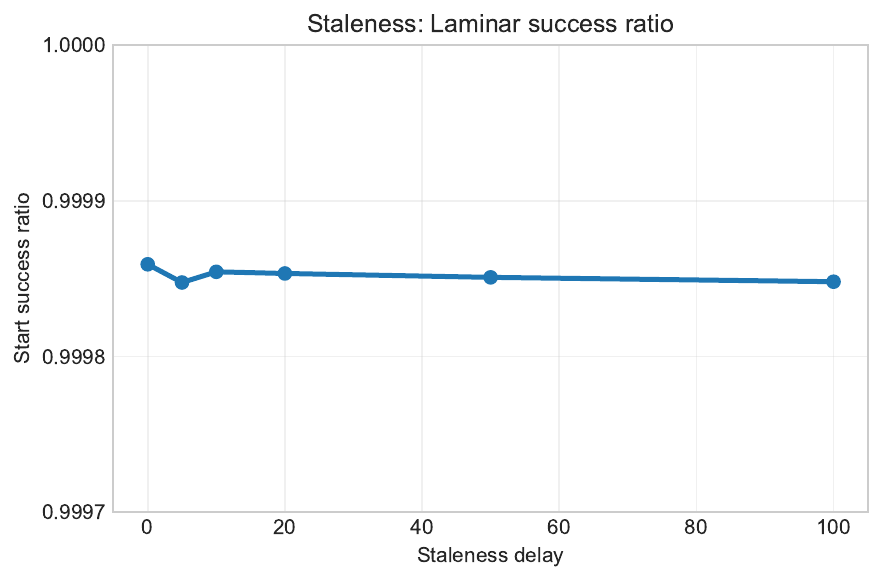}
  \end{minipage}
  \hfill
  \begin{minipage}[t]{0.48\columnwidth}
    \centering
    \includegraphics[width=\linewidth]{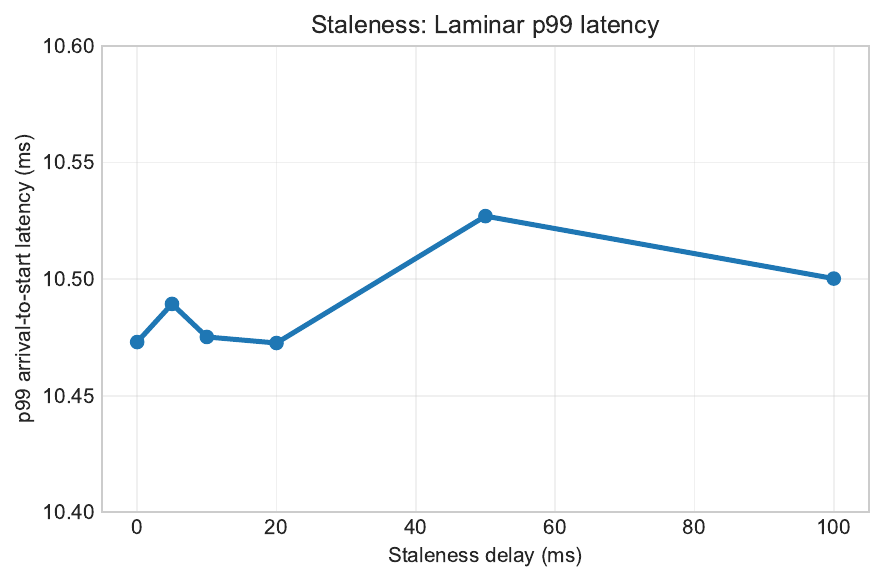}
  \end{minipage}
  \caption{Start success ratio (left) and p99 arrival-to-start
  latency (right) for Laminar as Z-HAF synchronization delay is
  swept from 0 to 100~ms at fixed $\rho=0.8$, 5{,}000-node
  cluster.}
  \label{fig:exp3}
\end{figure}

Figure~\ref{fig:exp3} shows that across all six delay settings,
p99 latency varies only from 10.47 to 10.53~ms and success ratio
stays within 99.98\%--99.99\%. This near-zero sensitivity to
delay up to 100~ms is not incidental: Z-HAF's short-project,
long-degrade rule absorbs brief outages through local state
projection, while node-local arbitration provides a final
physical rejection gate that prevents a stale optimistic
estimate from crossing into execution. Staleness therefore
inflates the probe dissipation rate at most, rather than
translating into latency amplification or execution-layer
errors.

\subsection{Mechanism Ablations (Exp4)}
\label{sec:exp4}

The stability observed in Exp1--3 rests on two specific
mechanisms. We isolate each with a controlled ablation,
disabling all other variable factors to remove confounds.

\begin{figure}[htbp]
  \centering
  \begin{minipage}[t]{0.48\columnwidth}
    \centering
    \includegraphics[width=\linewidth]{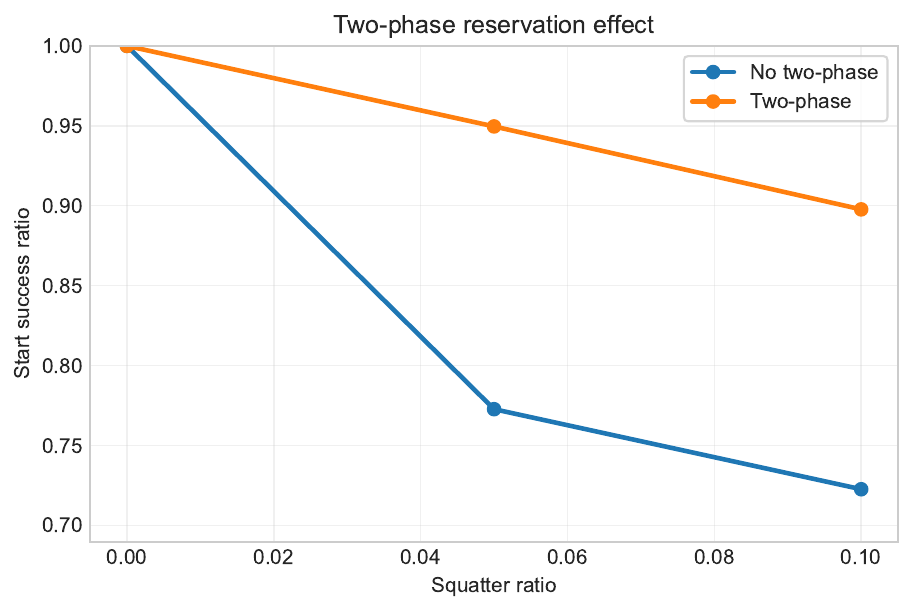}
  \end{minipage}
  \hfill
  \begin{minipage}[t]{0.48\columnwidth}
    \centering
    \includegraphics[width=\linewidth]{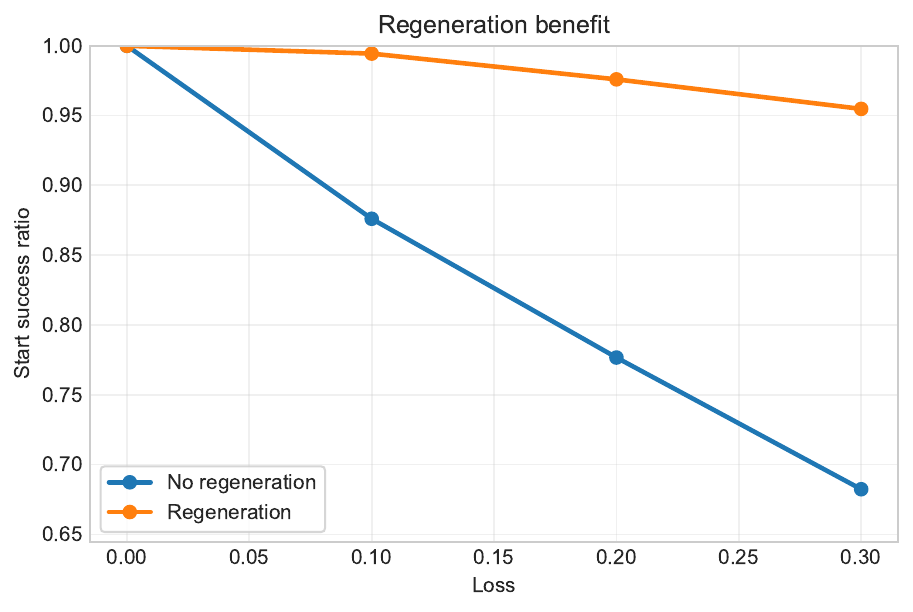}
  \end{minipage}
  \caption{(left)~Start success ratio with and without two-phase
  reservation at squatter ratios 0.05 and 0.10, $\rho=0.5$,
  DA regeneration disabled.
  (right)~Start success ratio with and without DA regeneration
  at probe loss rates 0.1, 0.2, and 0.3, $\rho=0.8$,
  two-phase reservation disabled.}
  \label{fig:exp4}
\end{figure}

\textbf{Two-phase reservation.}
We fix $\rho=0.5$, disable DA regeneration, and sweep squatter
ratio over $\{0.05, 0.10\}$ with two-phase reservation toggled
on and off over 5~s. A squatter wins arbitration but never completes
payload pull, occupying node-local state indefinitely.
Without two-phase reservation, success ratio drops to 77.27\%
and 72.27\%; enabling it recovers these to 94.95\% and 89.77\%,
confirming that TTL-bounded reservations with a frozen deposit
prevent squatters from silently exhausting arbitration capacity.

\textbf{DA regeneration.}
We fix $\rho=0.8$, disable two-phase reservation, and sweep
packet loss rate over $\{0.1, 0.2, 0.3\}$ with regeneration
toggled on and off over 30~s. Without regeneration, success ratio falls
to 87.60\%, 77.66\%, and 68.23\%; with regeneration, these
recover to 99.43\%, 97.59\%, and 95.47\%, absorbing network
loss as bounded control-plane dissipation rather than an
irreversible throughput drop.


\subsection{Runtime Survival with Airlock (Exp5)}
\label{sec:exp5-airlock}

To isolate Laminar's runtime-survival semantics, Exp5 compares two
otherwise identical Laminar configurations that differ only in whether
Airlock is enabled. The goal is not to maximize admission success in
isolation, but to test whether Laminar can replace opaque post-start
OOM destruction with an ordered, bounded survival policy once severe
node-local memory pressure emerges.

Unless otherwise stated, Exp5 uses Laminar on a 5{,}000-node cluster at
$\rho=0.8$ with zero packet loss over a 30\,s run. We retain the same
heterogeneous Zone configuration used in the main evaluation
(heterogeneous Zones enabled, target Zone size 256, and Zone-size
jitter 0.20), while disabling two-phase reservation and DA
regeneration so that the observed differences arise from Airlock
itself rather than from other recovery mechanisms. To induce sustained
runtime pressure, we enable dynamic memory perturbation with
runtime-pressure thresholds $(0.90, 0.80)$ for high/safe operation,
overclaim probability 0.3 with maximum overclaim factor 0.5, memory
drift $\kappa=0.10$, Gaussian memory noise $\sigma=0.1$, burst rate
0.02, burst scale 0.25, and a runtime tick of 1\,ms. 

\begin{figure}[htbp]
  \centering
  
  \includegraphics[width=\linewidth]{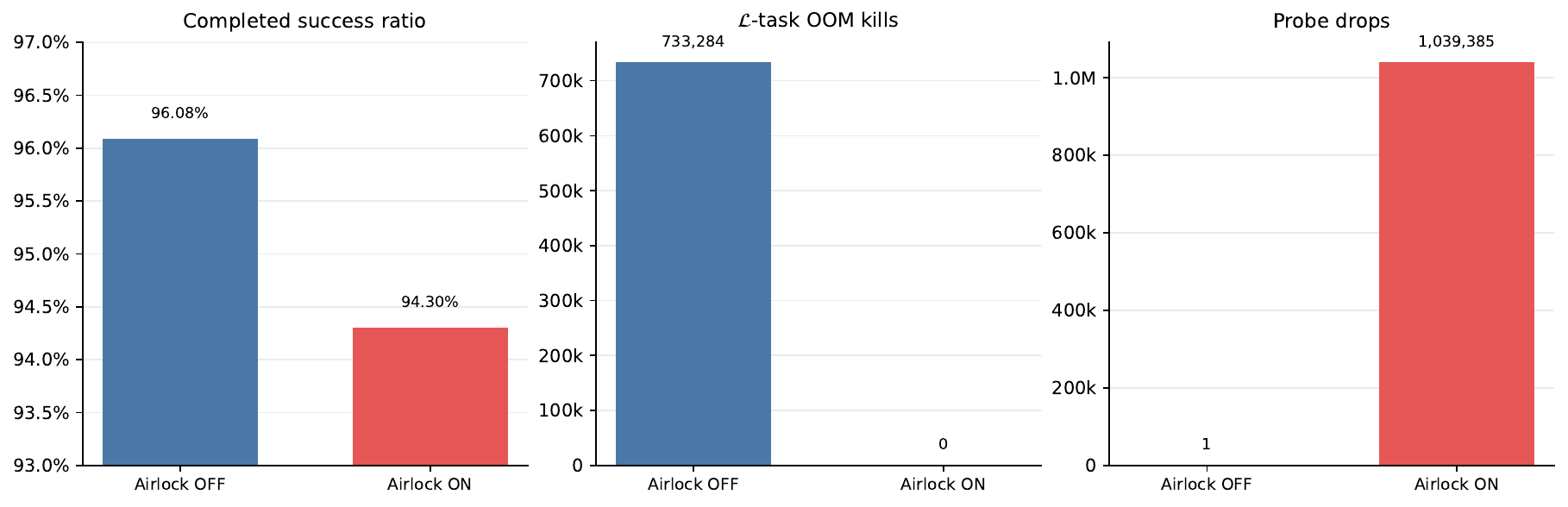}
  
  \vspace{0.5cm} 
  
  \includegraphics[width=\linewidth]{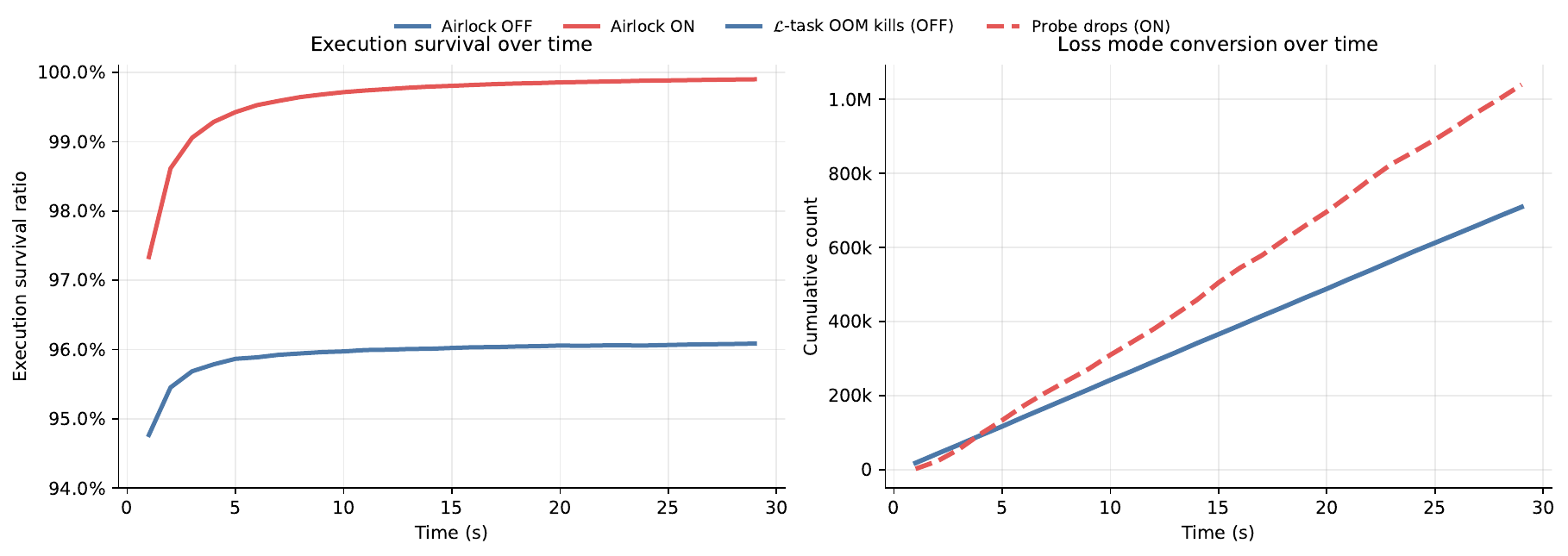}
  
  \caption{Exp5 runtime-survival behavior at 5{,}000 nodes and $\rho=0.8$.
  The top panel summarizes the end-of-run outcome, showing completed
  success ratio, $\mathcal{L}$-task OOM kills, and bounded probe dissipation with
  Airlock disabled and enabled. The bottom panel shows the time evolution
  of execution survival and loss-mode conversion. Without Airlock,
  $\mathcal{L}$-task OOM kills accumulate throughout the run and execution survival
  stabilizes near 96.1\%; with Airlock enabled, $\mathcal{L}$-task OOM kills remain
  at zero, execution survival approaches 99.9\%, and pressure is
  absorbed instead through bounded probe dissipation.}
  \label{fig:exp5_airlock}
\end{figure}

Figure~\ref{fig:exp5_airlock} shows that Airlock makes the system more useful under extreme pressure by preserving high-value $\mathcal{L}$-tasks and eliminating $\mathcal{L}$-task OOM kills, but it does so by converting more marginal low-priority tasks into earlier, bounded dissipation. Without
Airlock, Laminar attains a completed-success ratio of 96.08\%, but
incurs 733{,}284 $\mathcal{L}$-task OOM kills by the end of the run. With Airlock
enabled, $\mathcal{L}$-task OOM kills fall to zero, while the completed-success
ratio becomes 94.30\%. At the same time, bounded probe dissipation
rises sharply: cumulative probe drops increase from essentially zero
without Airlock to 1{,}039{,}385 with Airlock.

The time-series view clarifies how this transition unfolds. Without
Airlock, execution survival improves only slightly and then stabilizes
near 96.1\%, while $\mathcal{L}$-task OOM kills accumulate monotonically throughout
the run, reaching 709{,}269 by 29\,s. With Airlock enabled, $\mathcal{L}$-task OOM
kills remain at zero for the entire run, and execution survival rises
steadily toward 99.9\%. The price of this stronger runtime protection
is that pressure is absorbed earlier in the control path: probe drops
accumulate continuously and exceed 1.0 million by the end of the run.

These results are consistent with Laminar's intended semantics.
Airlock is not a free throughput gain; it is a runtime-survival layer
that inserts a deterministic decision process between local memory
pressure and task death. In this experiment, enabling Airlock causes
Laminar to reject or dissipate more work before execution fully
commits, but it simultaneously eliminates $\mathcal{L}$-task OOM destruction and
substantially raises execution survival. Exp5 therefore supports the
claim that Laminar extends scheduling responsibility beyond execution
start and converts uncontrolled runtime collapse into an ordered,
bounded degradation path.
\section{Related Work}
\label{sec:related}

\textbf{Centralized schedulers.}
Borg~\cite{borg}, Kubernetes~\cite{kubernetes}, Slurm~\cite{slurm},
Omega~\cite{omega}, Mesos~\cite{mesos}, YARN~\cite{yarn}, and
Firmament~\cite{firmament} maintain global resource views backed by
strongly consistent stores, shared-state optimistic concurrency, or
min-cost flow solvers. As analyzed in \S\ref{sec:motivation}, this
global visibility introduces serialized coordination that saturates
under high-arrival mixed workloads. Laminar routes via Zone-level
aggregates and closes placement at the node boundary without
distributed write locks. Furthermore, these architectures generally
end their semantic responsibility at admission, leaving runtime survival
to opaque kernel OOMs. Laminar introduces a node-local runtime-survival layer that provides a bounded path for suspension, in-situ recovery, secondary re-addressing, or reclamation without burdening the common-case hot path.

\textbf{Hierarchical schedulers.}
Hierarchical schedulers like Flux~\cite{flux} partition scheduling authority into recursive trees. Under high churn, concurrent sibling
decisions collide at shared leaves and require corrections to
propagate back up the hierarchy---the structure-bound bottleneck
in \S\ref{sec:motivation}. Laminar eliminates this path: contention
closes at a single node with no parent to traverse on failure. When
acute physical pressure occurs, it is absorbed by node-local data-plane
suspension rather than triggering structure-bound control-plane rollbacks.

\textbf{Decentralized schedulers.}
Apollo~\cite{apollo} uses opportunistic scheduling with optimistic
concurrency; Ray~\cite{ray} uses local-first placement with GCS
spillback. Both generate correction or redirection traffic under
high utilization---$\mathcal{O}(MN)$ RPCs in Ray's worst case.
Laminar makes failed search a local soft-state event: an
exhausted probe triggers Fast-Fail and dissipates in place.
Crucially, Laminar strictly isolates control and data planes during
post-landing eviction: the $\mathcal{O}(1)$ control-plane DA simply
re-addresses, while the heavyweight suspended state is pulled
asynchronously via the data plane, avoiding the global control-plane
RPC storms inherent in Ray's spillback.

\textbf{Probe-based and soft-state approaches.}
Power-of-two-choices~\cite{pow2} and Sparrow~\cite{sparrow} show
that small candidate sampling suffices for load balance. Laminar
builds on this but adds three absent elements: TTL-bounded
reservations that make waiting non-free; a patience budget that
bounds search depth; and DA regeneration that absorbs probe loss
as control-plane dissipation rather than application-visible
failure. Additionally, Laminar extends the probe's lifecycle into
a resident sentinel to anchor data-plane state pulls during secondary
reactivation.

\textbf{ML-specific schedulers.}
Tiresias~\cite{tiresias}, Pollux~\cite{pollux}, Themis~\cite{themis},
and Gavel~\cite{gavel} optimize JCT, GPU utilization, or fairness
via post-admission policy. Laminar is orthogonal: it addresses the
control-path cost to execution start and provides a deterministic
data-plane survival substrate (Airlock), delegating global epoch-level
management to these higher layers.
\section{Conclusion and Future Work}
\label{sec:conclusion}

In this paper, we identified the fragmented \emph{post-landing cluster ecology} as a distinct scheduling regime where Long tasks and fine-grained transient tasks must survive under severe physical constraints. We demonstrated that existing centralized, hierarchical, and decentralized paradigms reach structural bottlenecks in this regime because they couple resource discovery to heavyweight coordination and, crucially, terminate their semantic responsibility at execution start. This forces tenants to pay a "fear premium" to hedge against opaque host-level OOM heuristics.

Laminar replaces this fragile coupling with a probe-first, execute-later paradigm that formally extends its scheduling guarantee into runtime survival. By organizing the control path into macroscopic probabilistic routing, Zone-local projected state, and bounded node-local arbitration, Laminar achieves near-$\mathcal{O}(1)$ control-plane work on the hot path. More fundamentally, Laminar introduces a bipartite lifecycle for the Decentralized Agent (DA) and a node-local Airlock protocol. Together, they transform acute physical memory pressure from a catastrophic failure event into a deterministic, priority-ordered degradation path. Through in-situ temporal suspension, shared survival TTLs, and bounded secondary DA reactivation, Laminar structurally insulates high-priority workloads from blind kernel termination. This priority-ordered survival policy reduces the incentive to hoard excess capacity as a fear premium, helping the cluster operate closer to physical saturation without relying on protocol-level overcommitment.

Laminar is deliberately scoped to single-node placement and survival for residual workloads, leaving several compelling avenues for future work. First, the probe-first discipline currently treats the placement of upper-layer rigid-topology jobs as an exogenous given. A natural extension is to construct a bidirectional feedback loop where the aggregated fluidic fragmentation signals from Z-HAF are exposed to the macro-scheduler, allowing rigid-topology placement algorithms to minimize downstream ecological disruption. 

Second, Laminar's current runtime survival mechanisms rely on configured temporal bounds, such as the suspension threshold $T_{\text{susp}}$ and patience budgets. Future work will explore integrating lightweight, node-local reinforcement learning to dynamically tune these temporal buffers based on empirical host recovery rates. The primary architectural challenge in such an extension will be embedding these predictive heuristics without violating Laminar's core constraint: preserving the strict, lock-free $\mathcal{O}(1)$ execution path that allows the system to scale predictably under exascale arrival rates.

\bibliographystyle{IEEEtran}
\bibliography{refs}

\end{document}